\documentclass[fleqn,usenatbib]{mnras}
\usepackage[T1]{fontenc}
\usepackage{graphicx}
\usepackage{amsmath}
\usepackage{amssymb}
\usepackage{xspace}
\usepackage{newtxtext,newtxmath}
\usepackage{multirow}
\usepackage[section]{placeins}

\newcommand{\nv}{\hat{\bf n}}
\newcommand{\rsfr}{\rho_{\rm SFR}}
\newcommand{\bsfr}{\langle b\rho_{\rm SFR}\rangle}
\newcommand{\usfr}{M_{\odot}{\rm yr}^{-1}{\rm Mpc}^{-3}}
\newcommand{\lmax}{\ell_{\rm max}}
\newcommand{\nmt}{\texttt{NaMaster}\xspace}
\newcommand{\healpix}{\texttt{HEALPix}\xspace}
\newcommand{\ccl}{\texttt{CCL}\xspace}
\newcommand{\emcee}{\texttt{emcee}\xspace}
\newcommand{\planck}{{\sl Planck}\xspace}
\newcommand{\eboss}{eBOSS-QSO\xspace}
\newcommand{\dls}{DELS\xspace}

\DeclareRobustCommand{\VAN}[3]{#2}
\let\VANthebibliography\thebibliography
\def\thebibliography{\DeclareRobustCommand{\VAN}[3]{##3}\VANthebibliography}

\title[The star formation history from CIB correlations]{The star formation history in the last 10 billion years from CIB cross-correlations}

\author[Jego et al.]{Baptiste Jego$^{1,2}$\thanks{E-mail: baptiste.jego@ens-paris-saclay.fr},
Jaime Ruiz-Zapatero$^{2}$,
Carlos Garc\'ia-Garc\'ia$^{2}$,
Nick Koukoufilippas$^{2}$,
and David Alonso$^{2}$
\\
$^{1}$ENS Paris-Saclay, Gif-sur-Yvette, France\\
$^{2}$Department of Physics, University of Oxford, Denys Wilkinson Building, Keble Road, Oxford OX1 3RH, UK
}

\date{Accepted XXX. Received YYY; in original form ZZZ}

\pubyear{2022}

\begin{document}
\label{firstpage}
\pagerange{\pageref{firstpage}--\pageref{lastpage}}
\maketitle

\begin{abstract}
  The Cosmic Infrared Background (CIB) traces the emission of star-forming galaxies throughout all cosmic epochs. Breaking down the contribution from galaxies at different redshifts to the observed CIB maps would allow us to probe the history of star formation. In this paper, we cross-correlate maps of the CIB with galaxy samples covering the range $z\lesssim2$ to measure the bias-weighted star-formation rate (SFR) density $\bsfr$ as a function of time in a model independent way. This quantity is complementary to direct measurements of the SFR density $\rsfr$, giving a higher weight to more massive haloes, and thus provides additional information to constrain the physical properties of star formation. Using cross-correlations of the CIB with galaxies from the DESI Legacy Survey and the extended Baryon Oscillation Spectroscopic Survey, we obtain high signal-to-noise ratio measurements of $\bsfr$, which we then use to place constraints on halo-based models of the star-formation history. We fit halo-based SFR models to our data and compare the recovered $\rsfr$ with direct measurements of this quantity. We find a qualitatively good agreement between both independent datasets, although the details depend on the specific halo model assumed. This constitutes a useful robustness test for the physical interpretation of the CIB, and reinforces the role of CIB maps as valuable astrophysical probes of the large-scale structure. We report our measurements of $\bsfr$ as well as a thorough account of their statistical uncertainties, which can be used to constrain star formation models in combination with other data.
\end{abstract}

\begin{keywords}
    cosmology: large-scale structure of the Universe -- galaxies: star formation
\end{keywords}

\section{Introduction}\label{sec:intro}
  Dust in star-forming galaxies absorbs the stellar ultraviolet radiation from newly-formed, short-lived, massive stars, obscuring it, and re-emits it in the infrared (IR). The combined diffuse IR emission from all such sources thus constitutes a tracer of star formation \citep{1967ApJ...147..868P,2001ApJ...550....7K}, and is known as the Cosmic Infrared Background (CIB).
  
  Since its first detection by COBE \citep{1998ApJ...508..106D} (although see \citet{1996A&A...308L...5P}), the CIB has been both a nuisance and a blessing for cosmology and astrophysics. On the one hand, it is the most significant extra-galactic foreground affecting observations of the Cosmic Microwave Background (CMB) temperature fluctuations at high frequencies \citep{2011A&A...536A..18P,2013JCAP...07..025D}. Unlike other foregrounds, the fact that the IR spectral energy distribution (SED) is not universal across different galaxies and cosmic times, makes the CIB particularly difficult to separate in multi-frequency CMB observations. This complicates not only the cosmological analysis of small-scale CMB data, but also the recovery of unbiased maps of the CIB itself in the presence of Galactic dust contamination \citep{2014A&A...571A..30P,2016A&A...596A.109P,Lenz19}. Furthermore, since the CIB fluctuations trace the non-Gaussian large-scale structures that develop at low redshifts, it is an important systematic for CMB lensing analyses, which exploit the non-Gaussian modifications caused by the gravitational lensing of the otherwise Gaussian CMB fluctuations \citep{2014JCAP...03..024O,2014ApJ...786...13V,2021PhRvD.104l3514S,2021arXiv211100462D}. Given the central role that small-scale primary CMB data and CMB lensing will play in cosmology in the next decade \citep{2018PhRvD..97l3540S,2022MNRAS.509.5721F}, improving our understanding of the CIB, and developing better models that allow us to quantify its impact on these observations, is of paramount importance.
  
  On the other hand, the CIB is in itself a treasure trove for both astrophysics and cosmology. First, it is a high-sensitivity tracer of the star-formation rate (SFR) density (i.e. the total mass of stars created per unit time and volume) \citep{2006A&A...451..417D}. Understanding star formation across time and galaxy type is a central piece in the puzzle of galaxy formation and evolution \citep{1980FCPh....5..287T}. Through observations of the infrared luminosity function at high redshifts \citep{2013MNRAS.432...23G,2013A&A...553A.132M,2016MNRAS.456.1999M,2016MNRAS.461..458D}, a qualitative picture has emerged where star formation grows from the epoch of reionisation, peaks at around $z\sim2$, and then decreases as the gas that fuels it runs out \citep{2014ARA&A..52..415M}. However, collecting large samples of star-forming galaxies at high redshift is observationally challenging. Thus, since the CIB combines the emission of all IR sources since $z\sim6$, it encodes information about the full history of star formation, making it a useful complementary probe for these studies.
  
  Unfortunately, studies of the star formation history from maps of the CIB alone \citep[e.g.][]{2012MNRAS.421.2832S,2013ApJ...772...77V,2014A&A...571A..30P,Maniyar_2018,2021A&A...645A..40M} are hampered by its projected nature. Without additional information it is not possible to disentangle, in a model-independent manner, the contribution to the CIB from sources at different redshifts. Although cross-correlations with CMB lensing can be used to improve constraints on different CIB models \citep{2014A&A...571A..18P,Maniyar_2018,2020ApJ...901...34C,2021MNRAS.500.2250D,2021PhRvD.103j3515M}, they suffer from the same problem. This, however, can be solved through the use of {\sl galaxy clustering tomography} \citep{2019ApJ...877..150C}. Since the sources that give rise to the CIB trace the same large-scale structure as any other galaxy sample, the amplitude of the angular correlation between the CIB and a set of galaxies at a sufficiently narrow range of redshifts will be proportional to the contribution to the CIB map from sources at those redshifts. This technique has been widely utilised in various areas of cosmology and astrophysics to reconstruct the evolution, or the redshift dependence, of different quantities. These applications include recovering the mean gas pressure from maps of the thermal Sunyaev-Zel'dovich effect \citep{2019PhRvD.100f3519P,2020MNRAS.491.5464K,2020ApJ...902...56C,2102.07701}, the cosmic UV background \citep{2019ApJ...877..150C}, and the growth of structure \citep{2021JCAP...10..030G,2021JCAP...12..028K}, as well as the calibration of redshift distributions for imaging galaxy samples \citep[the so-called ``clustering redshifts'' method][]{2008ApJ...684...88N,2013arXiv1303.4722M,2017MNRAS.467.3576M}. Recently \cite{2022arXiv220401649Y} used cross-correlations between CIB maps and galaxies from the Kilo-Degree Survey (KiDS) to constrain a physical model of the CIB (see also \cite{2014A&A...570A..98S,2015MNRAS.449.4476W,2016ApJ...831...91C}).
  
  The aim of this paper is to use cross-correlations of the CIB with samples of galaxies at different redshifts to reconstruct the star formation history since $z\sim2$. Our approach will be somewhat different from that of \cite{2022arXiv220401649Y}. First, we will show that, from the amplitude of the galaxy auto-correlation and the cross-correlation with the CIB on large scales, robust and model-independent constraints can be placed on the evolution of the {\sl bias-weighted} star-formation rate density $\bsfr$. In this quantity, the contribution to the total SFR density from each source is weighted by the linear clustering bias of the dark matter halo it resides in. $\bsfr$ is therefore more sensitive to the contribution from galaxies in massive haloes than direct measurements of the SFR density $\rsfr$, and thus provides a complementary probe for the physics of star formation. The model-independent measurements of $\bsfr$ obtained can then be used to constrain the parameters of a halo-based SFR model. This will allow us to propagate our measurements onto constraints on $\rsfr$ that can be compared with direct measurements of this quantity from studies of the infrared luminosity function, thus providing a useful consistency test of the physical models used to describe star formation.

  This paper is structured as follows. Section \ref{sec:methods} lays out the theoretical framework, based on the halo model, that will allow us to measure $\bsfr$ as a function of redshift. The galaxy samples and CIB maps used in this analysis are then presented in Section \ref{sec:data}. Section \ref{sec:results} presents our model-independent measurements of $\bsfr$, quantifies their robustness, and propagates them onto constraints on $\rsfr$ and on two halo-based models for star formation. We then present our conclusions and outlook in Section \ref{sec:conc}.

\section{Methods}\label{sec:methods}
  \subsection{Theory}\label{ssec:methods.theo}
    \subsubsection{Projected probes}\label{ssec:methods.theo.proj}
      Let $u(\nv)$ be a cosmological field projected on the sphere, and assume that it is the radial projection of a 3D field $U$, i.e.
      \begin{equation}\label{eq:projected.anisotropy}
        u(\nv) = \int d\chi q_u(\chi)\,U(\chi\nv,z).
      \end{equation}
      Here $\chi$ is the comoving distance, $q_u(\chi)$ is the radial kernel associated with $u$, and $z$ is the redshift corresponding to the comoving distance $\chi$. The angular power spectrum of two such quantities, $u$ and $v$, $C_\ell^{uv}$, defined as the covariance between their harmonic-space coefficients, is then related to the power spectrum of the corresponding 3D quantities, $P_{UV}(k,z)$, via a similar radial projection:
      \begin{equation}\label{eq:angular.c_ell}
        C_\ell^{uv} = \int \frac{d\chi}{\chi^2} q_u(\chi)q_v(\chi) P_{UV}\left(k = \frac{\ell + 1/2}{\chi},z\right).
      \end{equation}
      Here we have used the Limber approximation \citep{1953ApJ...117..134L}, which is appropriate when the radial kernels are much broader than the typical correlation length of the fields (this is true for the quantities studied here).

    \subsubsection{Halo model}\label{sssec:methods.theo.hm}
      The halo model \citep{Seljak_2000, peacock2000halo, cooray2002halo} is a versatile formalism to model the 3D correlations of cosmological quantities. In this framework, the power spectrum $P_{UV}(k)$ can be written as the sum of two contributions, the one-halo and two-halo terms, $P^{1h}_{UV}(k)$ and $P^{2h}_{UV}(k)$, with
      \begin{align}\label{eq:1-halo}
        &P^{1h}_{UV}(k) \equiv \int_{}^{}dM\, n(M) \langle U(k,M)V(k,M)\rangle.\\\label{eq:2-halo}
        &P^{2h}_{UV}(k) \equiv \langle bU(k)\rangle \langle bV(k)\rangle P_{\rm lin}(k).
      \end{align}
      Here, $n(M)$ is the halo mass function (the comoving number density of haloes per mass interval), $U(k,M)$ is the profile of quantity $U$ around a halo of mass $M$ in Fourier space, $P_{\rm lin}(k)$ is the linear matter power spectrum, and $\langle bU(k)\rangle$ is
      \begin{equation}
        \langle bU(k)\rangle\equiv\int dM\,n(M)\,b_h(M)\,\langle U(k,M)\rangle.
      \end{equation}

      On large scales, both $P^{1h}_{UV}$ and $\langle bU\rangle$ tend to a constant. This can be seen easily by writing $U(k,M)$ in the limit $k\rightarrow 0$ (i.e. when $e^{-i{\bf k}\cdot{\bf x}} \rightarrow 1$):
      \begin{equation}
        U(k\rightarrow0,M)=\int d^3x\,U({\bf x}, M).
      \end{equation}
      Thus, on sufficiently large scales, the power spectrum can be approximated as
      \begin{equation}\label{eq:hm_simple}
        P_{UV}(k)=\langle bU\rangle\langle bV\rangle P_{\rm lin}(k)+N_{UV},
      \end{equation}
      where $\langle bU\rangle$, $\langle bV\rangle$ and $N_{UV}$ are now scale-independent quantities. Crucially, $\langle bU\rangle$, has a simple physical interpretation: it is the cosmic mean value of quantity $U$ weighted by the halo bias.

      Extracting information from scales smaller than those for which the approximation above is appropriate requires tackling several modelling challenges. First, an accurate model of the scale dependence of the halo profiles (i.e. the distribution of $U$ and $V$ around haloes of different masses) is necessary. Secondly, the 1-halo term depends on the statistical correlation of the two quantities under study, which in principle requires a model of not only the mean profiles, but also of their statistical correlation. Although these are often implicitly assumed to be independent (i.e. $\langle UV\rangle=\langle U\rangle\langle V\rangle$), deviations from this can be highly relevant and lead to incorrect conclusions (particularly given the high precision with which small-scale correlations can be measured). In turn, the 2-halo term requires no knowledge of the joint distribution of the quantities. Finally, the halo model is known to fail at the $\gtrsim10\%$ level on scales straddling the transition between the 1-halo- and 2-halo-dominated regimes. This is likely caused by the overly simplified assumption of linear biasing used in vanilla versions of the halo model to describe the clustering of dark matter haloes \citep{2021MNRAS.502.1401M,2021MNRAS.503.3095M}.

      Tackling these challenges is a complex problem that can be addressed through more sophisticated astrophysical modelling, often at the cost of introducing new nuisance parameters that must be marginalised over. The obvious benefit of this approach is that it enables the use of smaller scales, potentially achieving tighter constraints on the model parameters. This is indeed a common approach used in previous analyses of CIB auto- and cross-power spectra (e.g.: \citet{Maniyar_2018,2022arXiv220401649Y}).
      
      Here, we choose to avoid this additional complexity, and opt for a robust and model-independent measurement of the phenomenological parameters in Eq. \ref{eq:hm_simple}, restricting our analysis to large scales. Once the bias-weighted quantities $\langle bU\rangle$ have been measured, they can be used to constrain the underlying physical model describing them (in this case the star-formation rate history). All that remains is designing this model for the two tracers under consideration: the projected galaxy overdensity and the CIB emissivity.

    \subsubsection{CIB and star formation}\label{sssec:methods.theo.cib_sfr}
      The CIB intensity at a given observed frequency $\nu$ is related to the infrared emissivity (energy emitted per unit frequency, time and comoving volume in the rest frame of the emitter) $j_{\nu(1+z)}$ via:
      \begin{equation}\label{eq:cib.intensity}
        I_\nu(\nv) = \int d\chi \frac{j_{\nu(1+z)}(\chi\nv,z)}{4\pi(1+z)}.
      \end{equation}
      Given a model for the specific infrared luminosity of haloes of mass $M$, $L_{\nu_e}(M)$, the mean comoving emissivity becomes
      \begin{equation}\label{eq:emissivity.luminosity}
        \langle j_{\nu(1+z)}\rangle = \int_{}^{}dM n(M)\,L_{\nu(1+z)}(M),
      \end{equation}
      where $n(M)$ is the halo mass function. We now proceed as in \citet{Maniyar_2018,2021A&A...645A..40M}. The specific infrared luminosity can be written in terms of the observed source flux $S_\nu$
      \begin{equation}\label{eq:luminosity.flux}
        L_{\nu(1+z)}=4\pi\chi^2(1+z)S_\nu.
      \end{equation}
      At the same time, the total infrared luminosity $L_{\rm IR}$ (integrated over frequencies) is tightly correlated with the star-formation rate of the source:
      \begin{equation}\label{eq:luminosity.sfr}
        {\rm SFR}=K\,L_{\rm IR},
      \end{equation}
      where the $K=10^{-10}\,M_\odot\,{\rm yr}^{-1}L_\odot^{-1}$ is the calibration constant between the far infrared luminosity and star formation rate \citet{1998ARA&A..36..189K,2012ARA&A..50..531K} for a Chabrier initial mass function \citep{2003PASP..115..763C}, and $L_{\rm IR}$ is integrated over the range 8-1000$\mu$m. Combining Eqs. \ref{eq:emissivity.luminosity}, \ref{eq:luminosity.flux}, and \ref{eq:luminosity.sfr}, the CIB intensity can be written as
      \begin{equation}
        I_\nu(\nv)=\int d\chi\,\frac{\chi^2S_\nu^{\rm eff}(z)}{K}\,\rsfr(\chi\nv,z),
      \end{equation}
      where $\rsfr$ is the star-formation rate density, given by the contribution of all haloes of different masses. $S^{\rm eff}_\nu(z) \equiv S_\nu(z) / L_{\rm IR}$ is the mean flux of sources at redshift $z$ normalised to a total luminosity of $L_\odot$. We use the estimates of  $S^{\rm eff}_\nu(z)$ for the three \planck{} frequency channels derived by \citet{2013A&A...557A..66B,2015A&A...573A.113B,2017A&A...607A..89B}, and made available by \citet{2021A&A...645A..40M}\footnote{See \url{https://github.com/abhimaniyar/halomodel_cib_tsz_cibxtsz}.}.

      Comparing this with Eq. \ref{eq:projected.anisotropy}, the CIB intensity $I_\nu(\nv)$ can therefore be understood as a projected tracer of the star-formation rate density $\rsfr$, with a radial kernel given by
      \begin{equation}\label{eq:kernel_cib}
        q_\nu(\chi)=\frac{\chi^2S_\nu^{\rm eff}(z)}{K}.
      \end{equation}
      The associated $\langle bU\rangle$ is the mean star-formation rate density weighted by halo bias:
      \begin{equation}\label{eq:bsfr}
        \bsfr \equiv \int dM\,n(M)\,b_h(M)\,{\rm SFR}(M),
      \end{equation}
      where ${\rm SFR}(M)$ is the mean star-formation rate in haloes of mass $M$. The mean profile, in turn, would provide a measurement of the mean star-formation rate density
      \begin{equation}\label{eq:rhosfr}
        \rsfr=\int dM\,n(M)\,{\rm SFR}(M).
      \end{equation}

      When interpreting the measured values of $\bsfr$, we will make use of a halo-model parametrisation similar to that of \citet{2018MNRAS.477.1822M} and \cite{2021A&A...645A..40M} to describe the SFR-mass relation. The model splits the total SFR of a given halo into the contribution from central (${\rm SFR}_c$) and satellite galaxies (${\rm SFR}_s$):
      \begin{equation}
        {\rm SFR}(M,z)={\rm SFR}_c(M,z)+{\rm SFR}_s(M,z).
      \end{equation}
      As haloes accrete gas, part of it is transformed into stars. The model parametrises this process as:
      \begin{equation}
        {\rm SFR}_c(M,z)=\eta(M,z)\,{\rm BAR}(M,z),
      \end{equation}
      where ${\rm BAR}$ is the baryon accretion rate, and $\eta$ is the efficiency with which those baryons are transformed into stars. The BAR is modelled after \citet{2010MNRAS.406.2267F} as
      \begin{equation}
        {\rm BAR}(M,z)=\dot{M}_0\frac{\Omega_b}{\Omega_M}\,\left(\frac{M}{10^{12}M_\odot}\right)^{1.1}(1+1.11z)\frac{H(z)}{H_0},
      \end{equation}
      with $\dot{M}_0=46.1 M_\odot\,{\rm yr}^{-1}$.
      
      We will consider two parametrisations for the efficiency $\eta(M,z)$:
      \begin{enumerate}
        \item {\bf Model A:} the model of \citet{2021A&A...645A..40M}. In this case
          \begin{equation}
            \eta(M,z)=\eta_{\rm max}\exp\left[-\frac{(\log M_{\rm max}-\log M)^2}{2\sigma_M^2(M,z)}\right],
          \end{equation}
          where $M_{\rm max}$ is the halo mass of maximum efficiency $\eta_{\rm max}$. The logarithmic width $\sigma_M$, describing the range of masses with efficient star formation, depends on both mass and redshift as
          \begin{equation}
            \sigma_M(M,z)=\sigma_{M,0}-\tau\Theta(M-M_{\rm max})\,{\rm max}(0, z_c-z),
          \end{equation}
          where $\Theta$ is the Heaviside step function.
           
          Fixing $z_c=1.5$ as in \citet{2021A&A...645A..40M}, the model is described by 4 free parameters: $\eta_{\rm max}$, $\log_{10}M_{\rm max}$, $\sigma_{M,0}$, and $\tau$.
        \item {\bf Model B:} the parametrisation of \citet{2018MNRAS.477.1822M}. In this case:
          \begin{equation}
            \eta(M,z)=\frac{2\eta_*}{(M_1/M)^\beta+(M/M_1)^\gamma},
          \end{equation}
          where $\eta_*$, $m\equiv\log_{10}M_1/M_\odot$, $\beta$, and $\gamma$ are functions of redshift of the form:
          \begin{equation}
            x(z)=x_0+x_z\frac{z}{1+z}
          \end{equation}
          Fixing $(\beta_0,\beta_z,\gamma_0,\gamma_z)$ to the best-fit values found by \citet{2018MNRAS.477.1822M}, $(3.344,-2.079,0.966,0)$, the model is described by 4 free parameters: $\eta_0$, $\eta_z$, $m_0$, and $m_z$.
          
          In order to compare this model with Model A, note that the peak efficiency in this case is $\eta_{\rm max}=\eta_*f_\eta$, and it is attained at a mass $M_{\rm max}=M_1\,f_M$,
          where
          \begin{equation}
            f_M=\left(\frac{\beta}{\gamma}\right)^{\frac{1}{\beta+\gamma}},\hspace{6pt}
            f_\eta=\frac{2\gamma}{\beta+\gamma}f_M^\beta.
          \end{equation}
          For the values of $(\beta_0,\beta_z,\gamma_0,\gamma_z)$ used here, both factors are roughly constant and close to 1 ($f_M$ varies between 1.33 and 1.23, and $f_\eta$ varies between 1.17 and 1.04, in the range $0<z<4$). Thus, to a good approximation, $\eta_*$ and $M_1$ provide the peak efficiency and its associated halo mass.
      \end{enumerate}
      \begin{figure}
        \centering
        \includegraphics[width=0.48\textwidth]{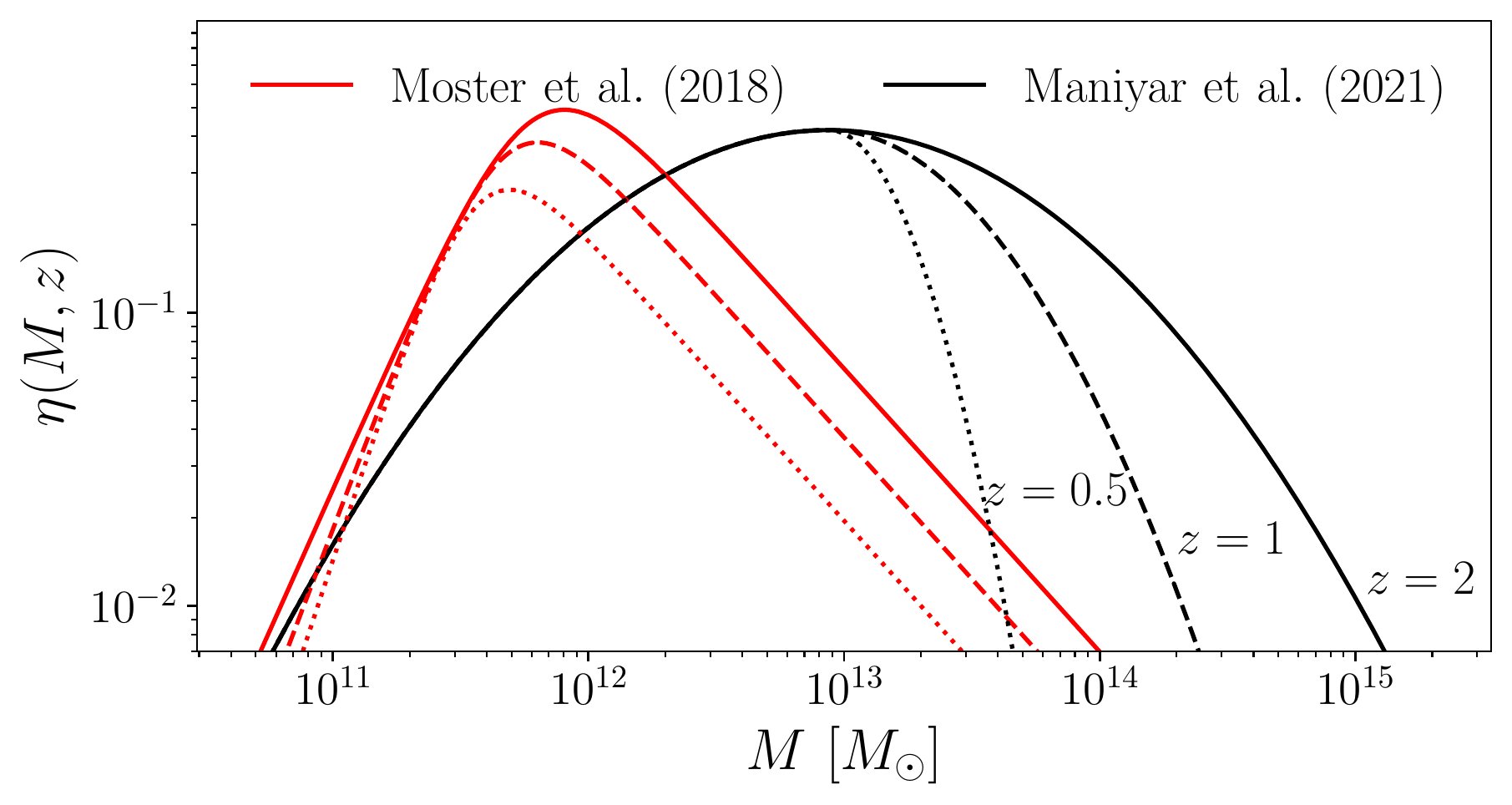}
        \caption{Mass-dependent star-formation efficiency at $z=0.5$ (dotted lines), $z=1$ (dashed lines) and $z=2$ (solid lines). Results are shown for the two halo-based models described in the text: Model A, from \citet{2021A&A...645A..40M}, in black, and Model B, from \citet{2018MNRAS.477.1822M}, in red. The curves shown correspond to the best-fit parameters found in the two papers ($\log_{10}M_{\rm max}=12.94$, $\eta_{\rm max}=0.42$, $\sigma_{M,0}=1.75$, $\tau=1.17$ for Model A, $m_0=11.34$, $m_z=0.692$, $\eta_0=0.005$, $\eta_z=0.689$ for Model B).}\label{fig:eta_models}
      \end{figure}

      Both models are qualitatively similar in that the star formation efficiency peaks at a particular halo mass, and decreases for lighter and heavier haloes. This is physically well motivated, based on the effect of low gravitational potentials and supernova feedback at low masses \citep{2003MNRAS.343..249S}, and an increased gas cooling time at high masses \citep{2005MNRAS.363....2K}. There are, however, important differences between them. On the one hand, the peak efficiency in Model A is constant as a function of time, and all time evolution is incorporated in the extent of the high-mass tail of $\eta(M,z)$. On the other, the peak efficiency is explicitly time-dependent in Model B, while the slope of the high-mass tail is time-independent. Furthermore, while the efficiency falls exponentially away from the peak mass in Model A, this drop is slower (power-law) in Model B. These differences are illustrated in Fig. \ref{fig:eta_models}, and they will allow us to explore the dependence of our final constraints on $\rsfr$ from our measurements of $\bsfr$ on the assumptions of the underlying model.

      Finally, the contribution from satellite galaxies will be calculated by adding the contribution of all subhaloes for each parent halo of mass $M$:
      \begin{equation}
        {\rm SFR}_s(M,z)=\int_{M_{\rm min}}^M dM_{\rm sub}\,\frac{dN}{dM_{\rm sub}}\,{\rm SFR}_{\rm sat}(M_{\rm sub},M,z),
      \end{equation}
      where $dN/dM_{\rm sub}$ is the subhalo mass function, calculated as in \cite{2010ApJ...719...88T}, and ${\rm SFR}_{\rm sat}(M_{\rm sub},M,z)$ is the SFR in a satellite galaxy with subhalo mass $M_{\rm sub}$ in a parent halo of mass $M$. As in \cite{2021A&A...645A..40M}, we model ${\rm SFR}_{\rm sat}$ as
      \begin{equation}\nonumber
        {\rm SFR}_{\rm sat}(M_{\rm sub},M,z)={\rm Min}\left[{\rm SFR}_c(M_{\rm sub},z),\frac{M_{\rm sub}}{M}{\rm SFR}_c(M,z)\right].
      \end{equation}

    \subsubsection{Projected galaxy clustering}\label{sssec:methods.theo.galaxy}
      The projected galaxy overdensity is related to its three-dimensional counterpart via:
      \begin{equation}\label{eq:kernel_gc}
        \delta_g(\nv) = \int d\chi H(z)p(z)\Delta_g(\chi\nv,z),
      \end{equation}
      where $p(z)$ is the galaxy redshift distribution for the galaxy sample under consideration, and $\Delta_g$ is its three-dimensional overdensity. Therefore, the kernel for this tracer is $q_g = H(z)\,p(z)$. The associated $\langle bU\rangle$ is the galaxy bias $b_g$.
      
      A popular halo model parametrisation for galaxy clustering is the halo occupation distribution (HOD) \citep{berlind2002halo}, based on a model for the number of galaxies as a function of halo mass $\bar{N}_g(M)$. In this framework, $b_g$ would simply be given by
      \begin{equation}
        b_g=\frac{\int dM\,n(M)\,b_h(M)\,\bar{N}_g(M)}{\int dM\,n(M)\,\bar{N}_g(M)}.
      \end{equation}
      We will fit directly for the value of $b_g$, and thus our treatment is insensitive to the details of the $\bar{N}_g(M)$ relation.

      The observed overdensity of galaxies is also affected by lensing magnification. Gravitational lensing by foreground structures induces a small perturbation in the observed galaxy positions and in their observed flux. This causes an additive contribution to $\delta_g$ of the form
      \begin{equation}
        \delta_g^M(\nv)=-\int d\chi\,(2-5s)\,q_L(\chi)\,\Delta_M(\chi\nv,z),
      \end{equation}
      where $\Delta_M$ is the 3D matter overdensity, $s=d\log_{10}N/dm$ is the logarithmic slope in the distribution of galaxy magnitudes at the limiting magnitude, and $q_L$ is the lensing kernel
      \begin{equation}
        q_L(\chi)\equiv\frac{3H_0^2\Omega_m}{2}(1+z)\chi\int_z^\infty\,dz'\,p(z')\,\frac{\chi(z')-\chi}{\chi(z')}.
      \end{equation}

      Due to the cumulative nature of the weak lensing kernel, if relevant, the contribution from magnification to the galaxy-CIB cross-correlation would bias the tomographic measurement of $\bsfr$ described in the next section. To verify that this contribution can indeed be ignored, we estimate the theoretical prediction with and without magnification for the cross-correlation between our highest eBOSS redshift bin (for which magnification is most relevant), and the CIB. We use $s=0.2$ for quasars \citep{2005ApJ...633..589S}. We find that magnification changes the cross-correlation only at the sub-percent level on the scales used here, and can therefore be ignored. Note that the impact of magnification on the \dls bins should be significantly smaller than for the quasar samples (see \citet{2010.00466}).

    \subsubsection{Tomography}\label{sssec:methods.theo.tomo}
      \begin{figure}
        \centering
        \includegraphics[width=0.48\textwidth]{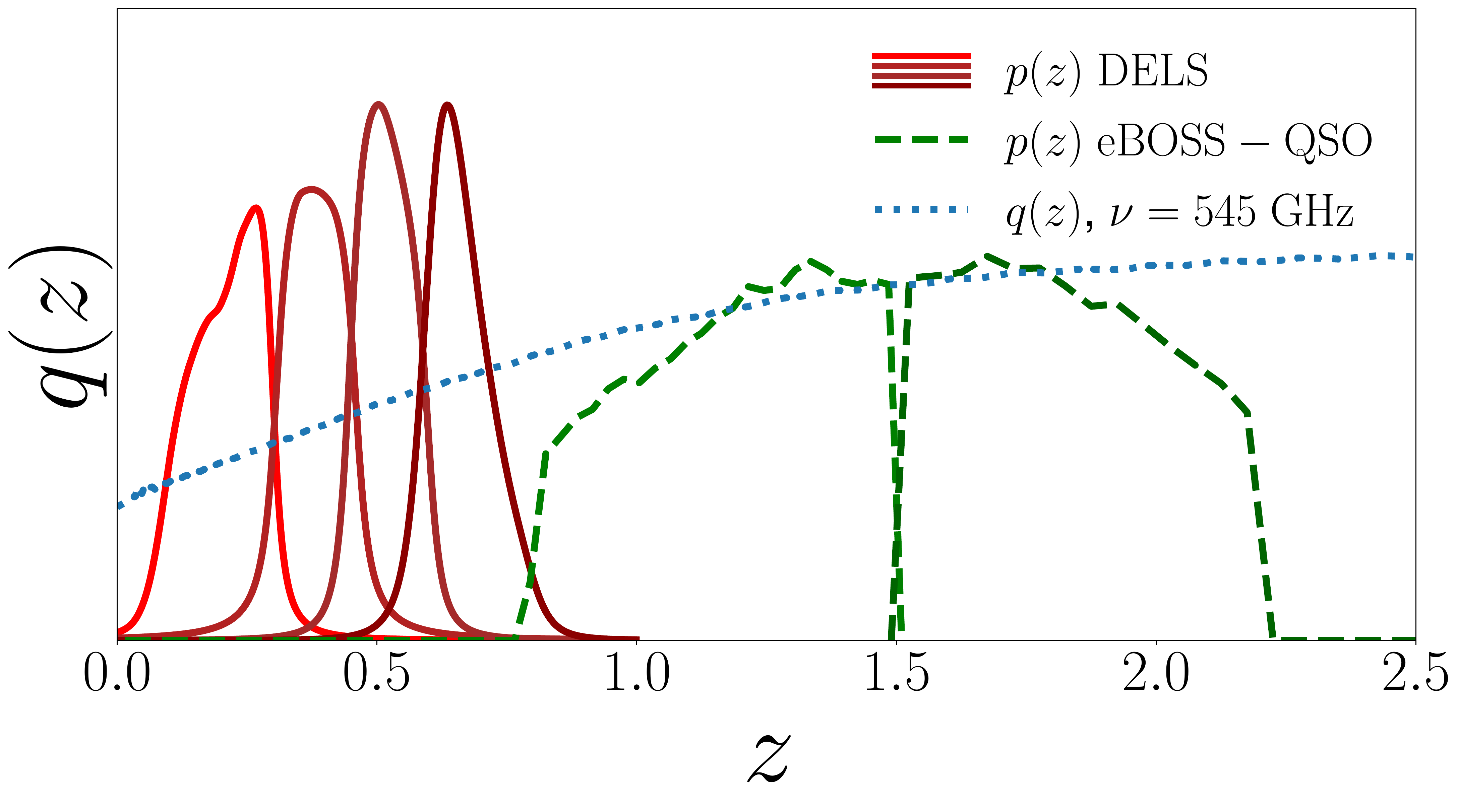}
        \caption{Redshift distributions of the $6$ redshift bins used in this work. The red lines show the distribution of the 4 \dls samples, while the dashed green lines show the two \eboss redshift bins. The dotted blue line shows the effective radial kernel for the CIB (Eq. \ref{eq:kernel_cib}) at 545 GHz. All curves have been rescaled by an arbitrary factor for visualisation purposes.}\label{fig:zbins}
      \end{figure}
      A crucial aspect of projected galaxy clustering is its local nature: galaxies at redshift $z$ trace structures at the same redshift. This makes it possible to directly measure the redshift evolution of astrophysical quantities by cross-correlating a given projected tracer (in our case CIB intensity maps) with galaxy samples at different redshifts, a procedure commonly called ``tomography'' \citep{2019ApJ...877..150C}.
      
      Consider the Limber equation (Eq. \ref{eq:angular.c_ell}), together with the simplified power spectrum model of Eq. \ref{eq:hm_simple}, and the radial kernels for the CIB (Eq. \ref{eq:kernel_cib}) and for galaxy clustering (Eq. \ref{eq:kernel_gc}). Assuming that the radial/redshift dependence of $b_g$, $\bsfr$, and $q_\nu(z)$ is slow compared with the width of the galaxy kernel $q_g(z)$, the auto-correlation of galaxies and their cross-correlation with the CIB can be approximated as
      \begin{align}\label{eq:cl_pheno}
        C_\ell^{gg}\simeq b_g^2\,M_\ell^{gg}+n_{gg}, \hspace{12pt}
        C_\ell^{g\nu}\simeq b_g\bsfr\,M_\ell^{g\nu}+n_{g\nu},
      \end{align}
      where $b_g$ and $\bsfr$ are evaluated at the mean redshift of the galaxy sample, and we have defined the power spectrum templates
      \begin{align}\label{eq:Ml}
        &M_\ell^{uv}\equiv \int\frac{d\chi}{\chi^2}q_u(\chi)q_v(\chi)P_M\left(\frac{\ell+1/2}{\chi},z\right),\\
        &n_{uv}\equiv\int\frac{d\chi}{\chi^2}q_u(\chi)q_v(\chi)N_{UV}(z),
      \end{align}
      where $P_M(k,z)$ is the matter power spectrum, and $N_{UV}$ was defined in Eq. \ref{eq:hm_simple}. Assuming a fixed cosmology, and an accurate knowledge of the radial kernels, $M^{gg}_\ell$ and $M^{g\nu}_\ell$ can be precomputed and treated as fixed templates. In that case, the galaxy auto-correlation effectively constrains the value of $b_g$, which then can be used to constrain $\bsfr$ at the redshift of the sample through the cross-correlation.

      Figure \ref{fig:zbins} shows the radial kernels of the different tracers used in this analysis. The CIB kernel for the $\nu=545\,{\rm GHz}$ \planck{} channel, shown as a blue dotted line (with arbitrary normalisation), varies very slowly in comparison with the redshift distributions of the six galaxy samples used, justifying our use of the approximate phenomenological model of Eq. \ref{eq:cl_pheno} to make tomographic measurements of $\bsfr$.
      
\begin{figure}
        \centering
        \includegraphics[width=0.48\textwidth]{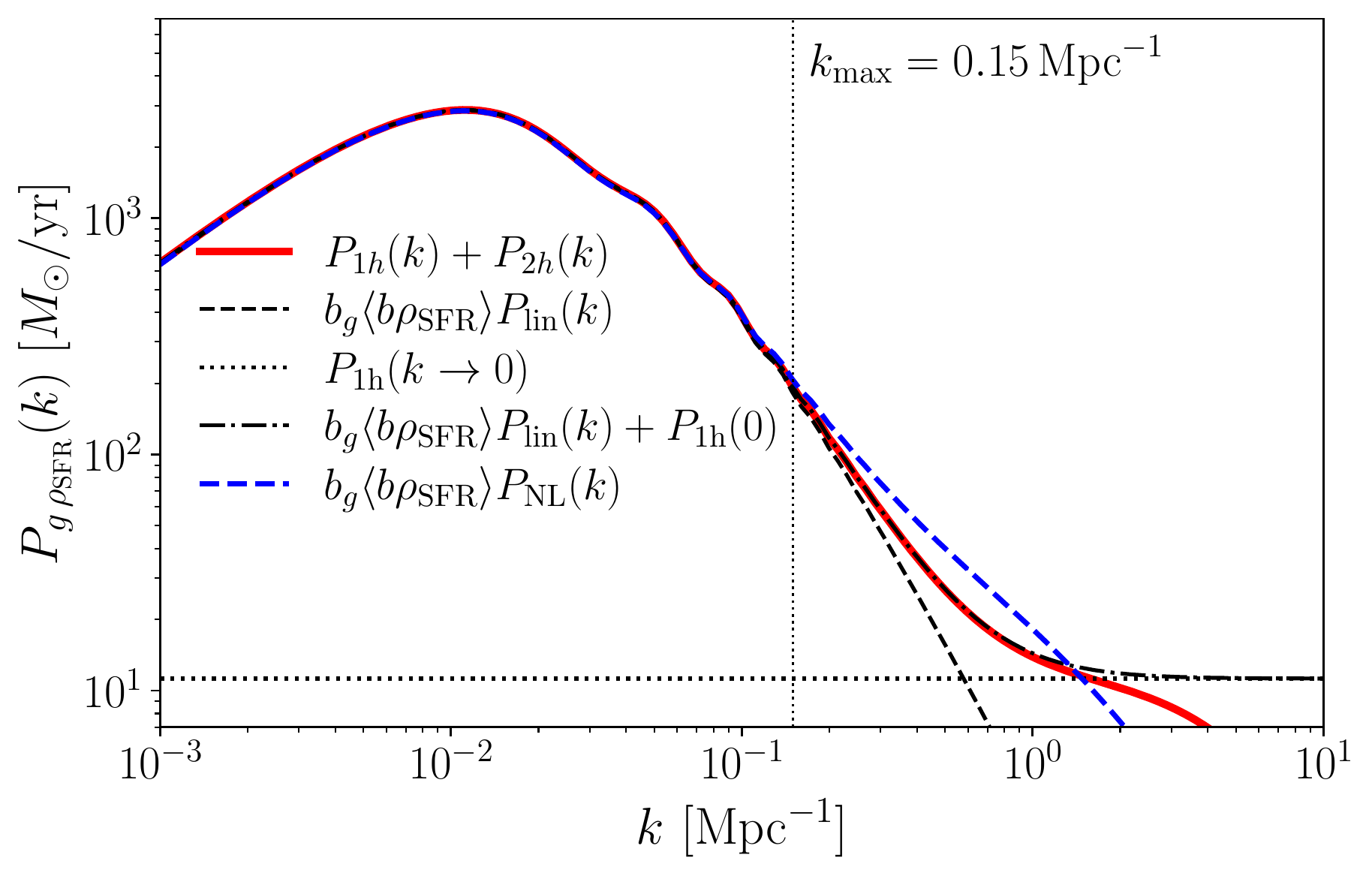}
        \caption{Halo model power spectrum at $z=0.8$ for the galaxy-$\rsfr$ cross-correlation. The solid red line shows the halo model prediction, while the black dash-dotted line shows the simplified phenomenological model of Eq.~\ref{eq:hm_simple}, using the $k\rightarrow0$ limit for the 1-halo term (black dotted line), and for the bias-weighted halo profiles in the 2-halo term (black dashed line). This approximation is accurate up to 2\% on scales $k \leq 1{\rm Mpc}^{-1}$, which justifies scale cuts used in Section \ref{ssec:results.cls}. The dashed blue line shows the simplified two-halo term using the non-linear matter power spectrum instead of the linear one.}
        \label{fig:spectra}
    \end{figure}

      Fig. \ref{fig:spectra} shows the cross-power spectrum between the galaxy overdensity and $\rsfr$ at $z\sim0.8$. The solid red line shows the prediction using the halo model from \citet{2021A&A...645A..40M}. This prediction can be recovered to $2\%$ accuracy up to $k=1\,{\rm Mpc}^{-1}$ using the simplified model of Eq. \ref{eq:cl_pheno}, given by the black dash-dotted line. This model is constructed by adding the linear matter power spectrum scaled by the product of the scale-independent biases associated with both quantities (dashed blue line), and the constant large-scale limit of the 1-halo term (black dotted line). However, in order to further improve the model and alleviate any residual theoretical uncertainties associated with the 1-2-halo transition regime, we use the {\sl non-linear} matter power spectrum, computed using the model of \citet{1208.2701}, scaled by the scale-independent biases (blue dashed line). This effective model is able to reproduce the halo model prediction to the same accuracy on scales $k<0.2\,{\rm Mpc}^{-1}$ without the additional shot noise terms. Therefore, we will by default use the non-linear matter power spectrum as $P_M$ when calculating the templates in Eq. \ref{eq:Ml}. We verified that the results shown in Section \ref{ssec:results.bsfr} change by less than $0.25\sigma$ if we use the linear power spectrum instead. Thus, the conservative choice of small-scale cut $k_{\rm max}=0.15\,{\rm Mpc}^{-1}$ used in our analysis (see Section \ref{sec:results}), marked as a vertical dotted line in the figure, should allow us to obtain unbiased constraints on $b_g$ and $\bsfr$ from the data. 

  \subsection{Angular power spectra and covariances}\label{ssec:methods.map2cl}
    We estimate auto- and cross-power spectra from all datasets in this study using the so-called pseudo-$C_\ell$ or MASTER algorithm as implemented in \nmt{} \citep{1809.09603}. The method is described briefly here, and further details can be found in \citet{astro-ph/0105302,1809.09603}.
    
    In all cases studied here (galaxy overdensity and CIB intensity maps), the observed field $\tilde{a}(\nv)$ is a masked version of the true underlying field $a(\nv)$, i.e.
    \begin{equation}
      \tilde{a}(\nv) = w(\nv) a(\nv),
    \end{equation}
    where $w$ is the mask. $w$ can simply be a binary map defining the footprint over which $a$ has been observed, but in general it can be thought of as a local weight that can be designed to maximise the signal-to-noise ratio of the resulting power spectrum estimator. Consider two such fields, $\tilde{a}$ and $\tilde{b}$, with masks $v$ and $w$ respectively. Their pseudo-$C_\ell$ is defined as $\tilde{C}^{ab}_\ell\equiv\sum_{m=-\ell}^\ell \tilde{a}_{\ell m}\tilde{b}^*_{\ell m}/(2\ell+1)$, where $\tilde{a}_{\ell m}$ are the harmonic coefficients of $\tilde{a}$. Due to the convolution theorem, the $\tilde{a}_{\ell m}$ are a convolution of the harmonic coefficients of the true map and the mask. This propagates to the power spectra, so that the expectation value of $\tilde{C}_\ell^{ab}$ is
    \begin{equation}\label{eq:pcl_mcm}
        \langle\tilde{C}^{ab}_\ell\rangle=\sum_{\ell'}M^{vw}_{\ell\ell'}C^{ab}_{\ell'}.
    \end{equation} 
    Here, $M^{vw}_{\ell\ell'}$ is the mode-coupling matrix (MCM), which depends only on the masks $v$ and $w$, and can be calculated analytically with relative ease. In order to estimate the true underlying power spectra from the observations we then need to invert the MCM. Since the MCM is not generally invertible, the pseudo-$C_\ell$ estimator often involves binning $\tilde{C}^{ab}_\ell$ into bandpowers before inverting the binned MCM. Here, we used the same bandpowers used in \citet{2021JCAP...10..030G}. Linear $\ell$ bins with $\Delta\ell=30$ were used up to $\ell=240$, and logarithmic bins were used at higher $\ell$ with $\Delta\log_{10}\ell=0.055$.
    
    The covariance matrix of angular power spectra is sourced by three contributions: a disconnected (or ``Gaussian'') part, a super-sample covariance (SSC) due to modes larger than the survey footprint, and a connected non-Gaussian (cNG) part caused by the intrinsic non-Gaussianity of the tracers \citep{2013PhRvD..87l3504T}. On the large scales used here the Gaussian component dominates, and we neglect the SSC and cNG contributions. We use an approximate analytical method to compute the Gaussian covariance matrix accounting for the effects of survey geometry, the so-called ``narrow-kernel approximation'' (NKA) described in \citet{1906.11765}, and implemented in \nmt. The NKA reduces the scaling of the exact calculation from  $O(\lmax^6)$ to a more tractable $O(\lmax^3)$ by assuming that the MCM is narrowly peaked at the diagonal compared with the slow scale dependence of the true power spectrum.

    Schematically, the covariance between two pseudo-$C_\ell$s can be approximated as
    \begin{equation}\nonumber
        {\rm Cov}(\tilde{C}^{ab}_\ell,\tilde{C}^{fg}_{\ell'})= {\cal C}^{ag}_{(\ell,\ell')}{\cal C}^{bf}_{(\ell,\ell')} \Xi_{\ell\ell'}(w^aw^g,w^bw^f)+(a\leftrightarrow b)\,
    \end{equation}
    where $\Xi_{\ell\ell'}(w^aw^g,w^bw^f)$ are coupling coefficients similar to those involved in the calculation of $M^{vw}_{\ell\ell'}$, but now dependent on the products of pairs of masks (labelled $w^{a/b/f/g}$ here). The modified spectra ${\cal C}^{ab}_\ell$ are related to the true underlying power spectra via \citep{2010.09717}
    \begin{equation}\label{eq:clcov}
        {\cal C}^{ab}_\ell\equiv \frac{\langle\tilde{C}^{ab}_\ell\rangle}{\langle w^a\,w^b\rangle_{\rm pix}},
    \end{equation}
    where $\langle\cdots\rangle_{\rm pix}$ denotes averaging over all pixels in the sky (as opposed to ensemble averaging). In general $\langle\tilde{C}^{ab}_\ell\rangle$ can be computed from a theoretical model and convolved with the MCM. However, since the power spectra estimated here have a high signal-to-noise ratio, we simply use the estimated pseudo-$C_\ell$s in the formula above (i.e. replace $\langle \tilde{C}_\ell^{ab}\rangle\rightarrow \tilde{C}_\ell^{ab}$). In addition to simplicity, this has the added benefit of automatically accounting for the noise contribution to the power spectrum with no additional modelling.

    This estimate of the covariance matrix was validated against jackknife resampling. As described in Section \ref{ssec:results.cls}, both estimates are in reasonable agreement, although a small modification was applied to the final covariance matrix elements involving two different CIB frequency maps to accurately account for their map-level correlation.

  \subsection{Likelihood}\label{ssec:methods.like}
    We derive parameter constraints from the power spectrum data in two different ways. First, as described in Section \ref{ssec:results.bsfr}, we derive model-independent tomographic measurements of $\bsfr$ from the galaxy auto-correlations and their cross-correlation with the CIB. We will then use these measurements, as described in Section \ref{ssec:results.const_hm}, to place constraints on the free parameters of the halo models for the SFR described in Section \ref{sssec:methods.theo.cib_sfr}. In both cases, we will make use of Gaussian likelihoods of the form
    \begin{equation}\label{eq:like}
      -2\log p({\bf d}|\Theta)=({\bf d}-{\bf m}(\Theta))^T{\sf C}_{\bf d}^{-1}({\bf d}-{\bf m}(\Theta))+K,
    \end{equation}
    where ${\bf d}$ is the data vector (power spectra or measurements of $\bsfr$), ${\sf C}$ is the covariance of ${\bf d}$, ${\bf m}$ is the theoretical model for ${\bf d}$, and $\Theta$ denotes the set of parameters used to define ${\bf m}$. The posterior parameter distribution is then given by the product of this likelihood and the parameter priors, which we discuss in Section \ref{ssec:results.bsfr}. In all cases we will use the affine-invariant Markov-Chain Monte-Carlo (MCMC) algorithm implemented in \emcee  \citep{emcee}.
      
    The assumption of a Gaussian likelihood for the power spectra stems from the central limit theorem: for sufficiently high $\ell$, the estimate of $C_\ell$ involves averaging over a large number of independent $m$ modes, and thus the distribution of $C_\ell$ values tends to a Gaussian. This is valid on the scales used in this analysis \cite{2008PhRvD..77j3013H}. The validity of the Gaussian approximation for the likelihood of the measured $\bsfr$ values will be demonstrated in Section \ref{ssec:results.bsfr}. 

    All cosmological theory predictions were computed using the Core Cosmology Library (\ccl, \cite{1812.05995}\footnote{The source code can be found at \url{https://github.com/LSSTDESC/CCL}}). Cosmological parameters were fixed to the \planck{} best-fit values $(\Omega_c,\Omega_b,h,n_s,\sigma_8)=(0.261, 0.049, 0.677, 0.9665, 0.8102)$. When using the halo model, we use the halo mass function parametrisation of \citet{Tinker_2008}, and the halo bias of \citet{2010ApJ...724..878T}. Halo masses are defined using a spherical overdensity $\Delta=200$ times the critical density.

\section{Data}\label{sec:data}
  \begin{figure}
    \centering
    \includegraphics[width=0.5\textwidth]{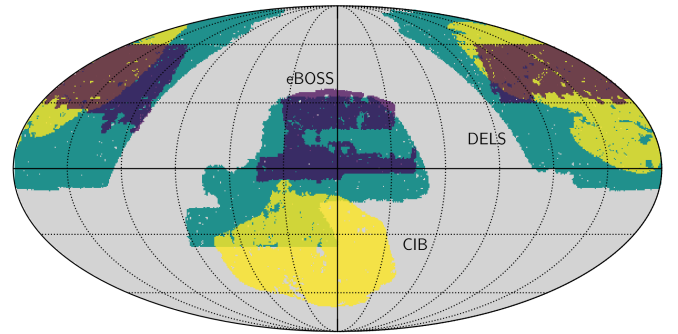}
    \caption{Sky footprint of the CIB and galaxy surveys used in this analysis in Equatorial coordinates.}
    \label{fig:footprint}
  \end{figure}
  \subsection{Galaxy clustering}\label{ssec:data.gc}
    \begin{table}
      \centering
      \def\arraystretch{1.2}
      \begin{tabular}{|cccc|}
        \hline
        Bin &  Redshift range & Mean redshift & Density (${\rm deg}^{-2}$)\\  
        \hline 
        \multicolumn{4}{c|}{\textbf{\dls}} \\
        1   & [$0.10, 0.30$) & $0.21$ & 808\\
        2   & [$0.30, 0.15$) & $0.37$ & 651\\
        3   & [$0.45, 0.60$) & $0.50$ & 760\\
        4   & [$0.60, 0.80$] & $0.63$ & 409\\
        \hline
        \multicolumn{4}{c|}{\textbf{eBOSS}} \\
        5   & [$0.80, 1.50$) & $1.12$ & 34\\ 
        6   & [$1.50, 2.20$] & $1.87$ & 35\\
        \hline
      \end{tabular}
      \caption{Mean redshift values and approximate redshift ranges for the $6$ redshift bins shown in Figure \ref{fig:zbins}. For \dls, the bin edges are in photo-$z$ space, while the mean redshift corresponds to the mean of the true redshift distribution. The 4th column shows the angular number density of each sample.}\label{tab:zbins}
    \end{table} %
    In order to recover the star formation history in the range $0\lesssim z\lesssim 2$ we make use of photometric galaxy samples from the DESI Legacy Imaging Survey \citep[\dls]{1804.08657}, and quasar samples (QSOs) from the extended Baryon Oscillation Spectroscopic Survey (eBOSS), derived the from Sloan Digital Sky Survey Data Release 16 (DR16).
    
    At low redshifts, we use the photometric sub-sample of DELS selected by \citet{2010.00466}\footnote{The data can be found at \url{https://gitlab.com/qianjunhang/desi-legacy-survey-cross-correlations}. We thank the authors for the remarkable care with which these data were made publicly available.}. \dls combines photometry from three different telescopes: the DECam Legacy Survey \citep{1504.02900,2016AAS...22831701B} at declinations ${\rm DEC}<33^{\circ}$, the Mayall $z$-band Legacy Survey \citep{2016AAS...22831702S}, and the Beijing-Arizona Sky Survey \citep{1908.07099}. The final survey covers 17739 deg$^2$. In \citet{2010.00466}, each galaxy was assigned a photometric redshift based on a multi-dimensional matching in colour space with a set of spectroscopic samples. We use these redshifts to separate the sample into the four tomographic bins used in \citet{2010.00466}, following the same procedure described there to model the redshift distributions of the resulting samples. Moreover, we impose low-declination cut of ${\rm DEC}>-36^\circ$, discarding a region exhibiting significant variations in completeness.

    At higher redshifts we use the homogeneous eBOSS quasar sample used for the cosmological power spectrum analysis of \citet{2007.08999,2007.08998}, and presented in \citet{2007.09000}. The catalogue comprises $343708$ objects with measured redshifts in the range $0.8 \leq z \leq 2.2$, covering over $4800$ deg$^{2}$. We combine the measurements from the North and South Galactic Caps into one single catalogue that we split into two different bins with redshifts above and below $z=1.5$ respectively. The redshift distribution of each bin is estimated directly from the data as a histogram of the measured spectroscopic redshifts.

    We construct maps of the galaxy overdensity from both samples using the methods described in \citet{2021JCAP...10..030G}. These include the construction of galaxy masks, the use of random catalogues for eBOSS, the correction of large-scale systematic fluctuations for \dls, and the estimation of the noise bias for both samples. Details can be found in Section 3 of \citet{2021JCAP...10..030G}. The impact of large-scale systematics in the galaxy auto-correlation is minimal given the scale cuts used for this analysis (see Section \ref{ssec:results.bsfr}). Likewise, our analysis is not sensitive to inaccuracies in the estimation of the shot-noise contribution, since we fully marginalise over a noise-like white component ($n_{xy}$ in Eq. \ref{eq:cl_pheno}) for all power spectra analysed. All maps are constructed and processed using \healpix{} \citep{2005ApJ...622..759G}, using a resolution parameter $N_{\rm side}=2048$, corresponding to a pixel size of $\sim1.7'$. Our final galaxy sample thus comprises 6 different redshift bins, shown in Fig. \ref{fig:zbins}, and described in Table \ref{tab:zbins}.

  \subsection{CIB maps}\label{ssec:data.CIBLenz}
    We use the full mission CIB maps and associated sky masks constructed by \citet{Lenz19}. The maps were created from the \planck{} 353, 545 and 857 GHz intensity maps. Galactic neutral hydrogen data from the HI4PI Survey \citep{bekhti2016hi4pi} were used to create template maps of galactic dust. These templates were then used to subtract potential galactic dust contamination in the maps. 

    We apply the sky mask constructed by \citet{Lenz19}, based on the $20$\% \planck{} Galactic plane mask, removing small-scale Galactic filamentary structure via HI column densities above $N_{\rm HI} > 2.5 \times 10^{20}\,{\rm cm}^{-2}$. The combined mask was then apodised using a  $15'$ FWMH kernel.

    When estimating the galaxy-CIB cross-correlations, we de-convolve the effective window function provided with the data products of \citet{Lenz19}. This accounts for the Planck Reduced Instrument Model (RIMO) beam, the window function of the SMICA CMB map, and the pixel window function for $N_{\rm side}=2048$ (see section 2.2 of \citet{Lenz19} for further details).

    As discussed in \citet{Lenz19}, the use of small sky patches to remove the Galactic dust component from the Planck maps leads to an over-subtraction of the underlying CIB fluctuations that results in a lack of power on large scales. The effect is limited to scales $\ell\lesssim70$, and therefore we will omit all such scales from our analysis. We applied color corrections as described in \cite{2014A&A...571A..30P}.

    The footprints of the three datasets used here are shown in Fig. \ref{fig:footprint}. The CIB maps overlap with the DELS and eBOSS footprints over $\sim 13\%$ and $\sim 6\%$ of the sky respectively.

\section{Results}\label{sec:results}
  \subsection{Power spectra and covariance}\label{ssec:results.cls}
    \begin{figure*}
        \centering
        \includegraphics[width=0.85\textwidth]{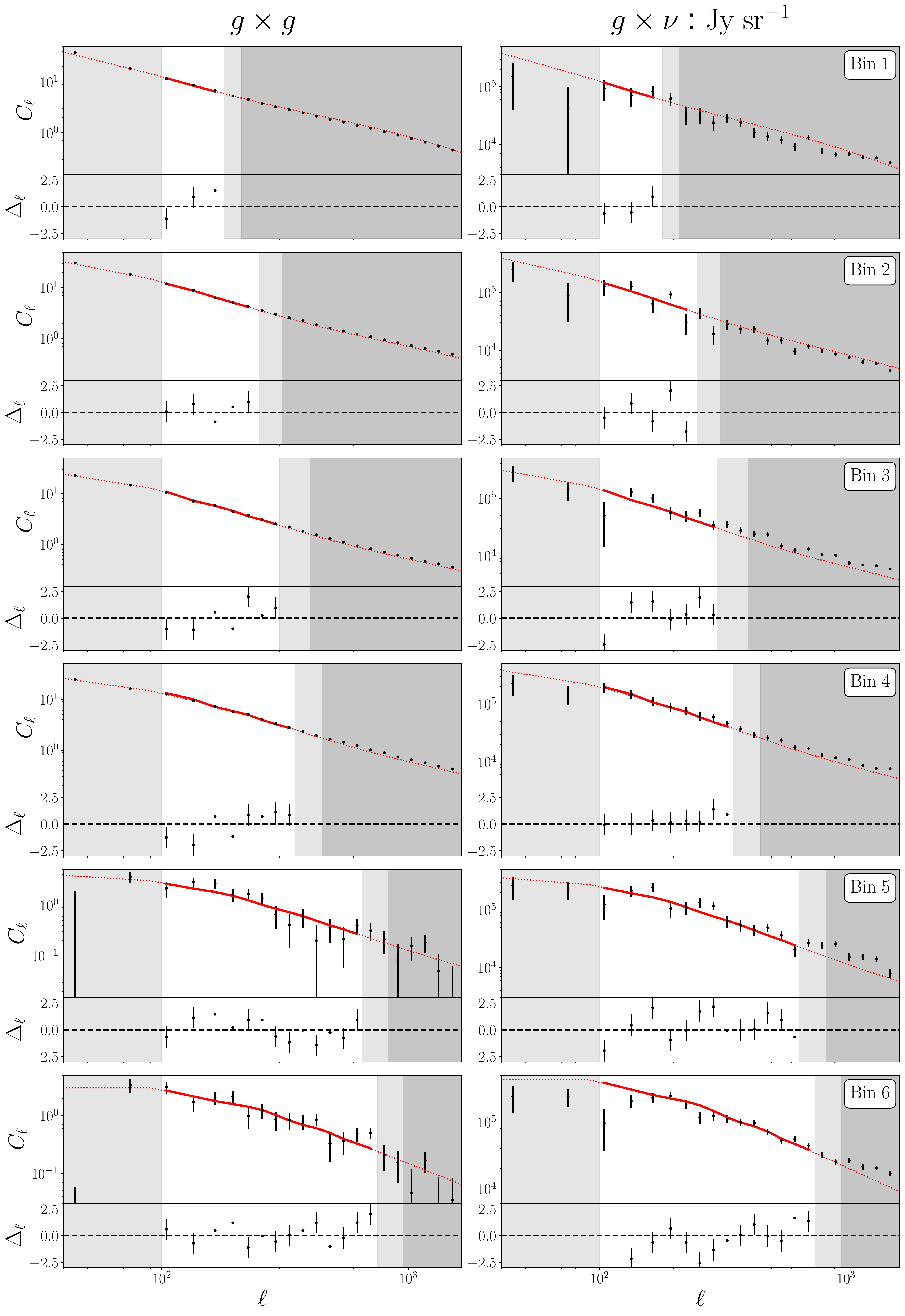}
        \caption{Galaxy auto-correlation spectra (left column) and galaxy-CIB cross-correlation spectra (right column) for the six tomographic bins presented in Fig. \ref{fig:zbins}. Each panel shows the measurements and their uncertainties (black points with error bars), and their best-fit predictions (red lines), which are extrapolated outside of the range of scales used in our analysis. Our fiducial scale cuts are marked by the light grey vertical bands, while the dark grey bands mark the alternative, less conservative scale cuts corresponding to $k_{\rm max} = 0.2\,{\rm Mpc}^{-1}$. The bottom part of each panel shows the difference between the data and the best-fist curve normalised by the $1\sigma$ uncertainties.}
        \label{fig:fitscuts}
    \end{figure*} %
    \begin{figure}
      \centering
      \includegraphics[width=0.48\textwidth]{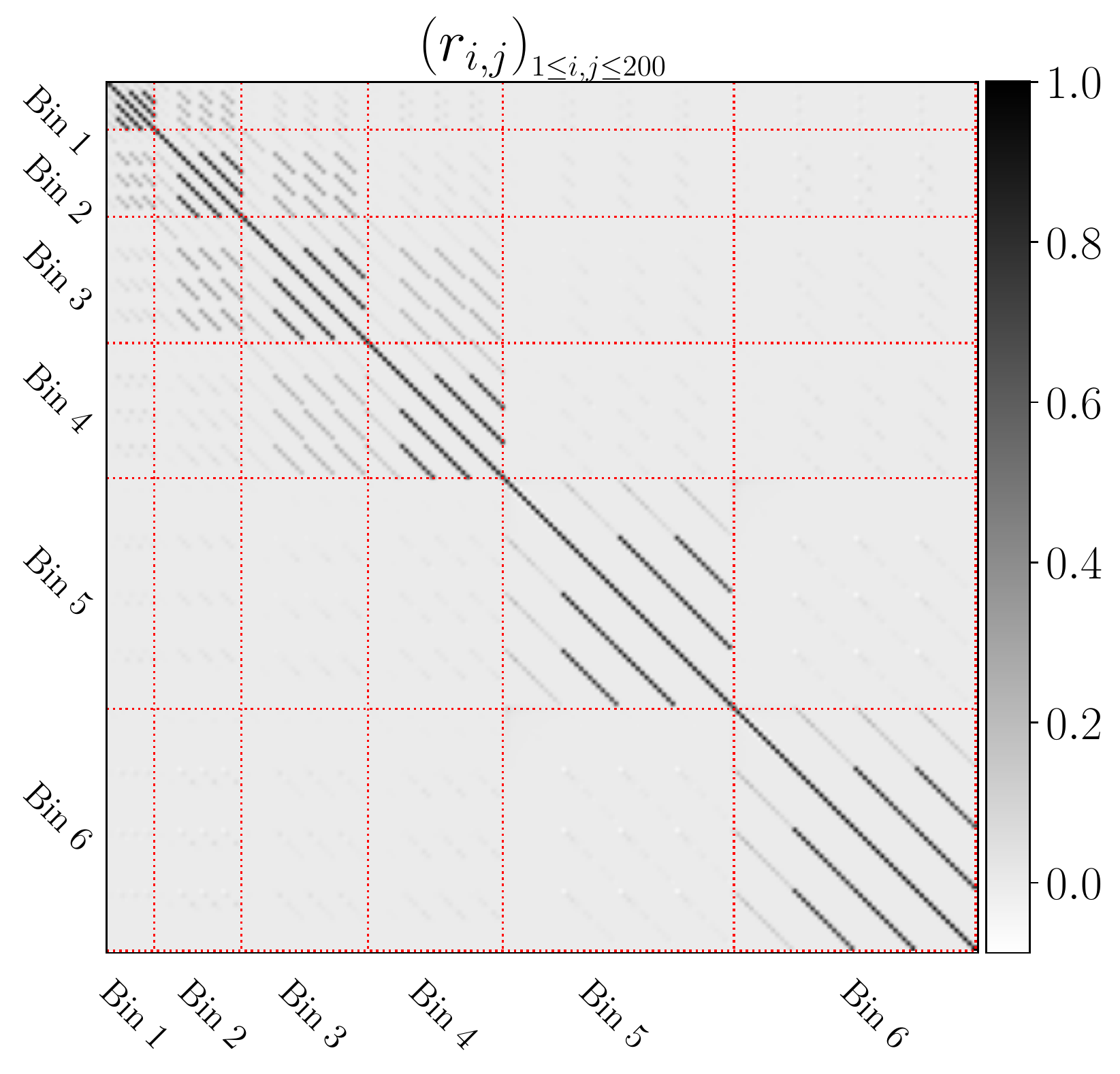}
      \caption{Correlation matrix $(r_{i,j})_{_{1\leq i,j \leq 200}}$ of the total data vector. The six blocks along the diagonal correspond to the different redshift bins used in the analysis. Within these blocks, the first sub-block contains the covariance of the galaxy auto-correlation, while the other blocks correspond to the different cross-correlations with the CIB. The different CIB frequency channels are strongly ($\sim95\%$) correlated.} \label{fig:covmat}
    \end{figure} %
    \begin{figure}
      \centering
      \includegraphics[width=0.48\textwidth]{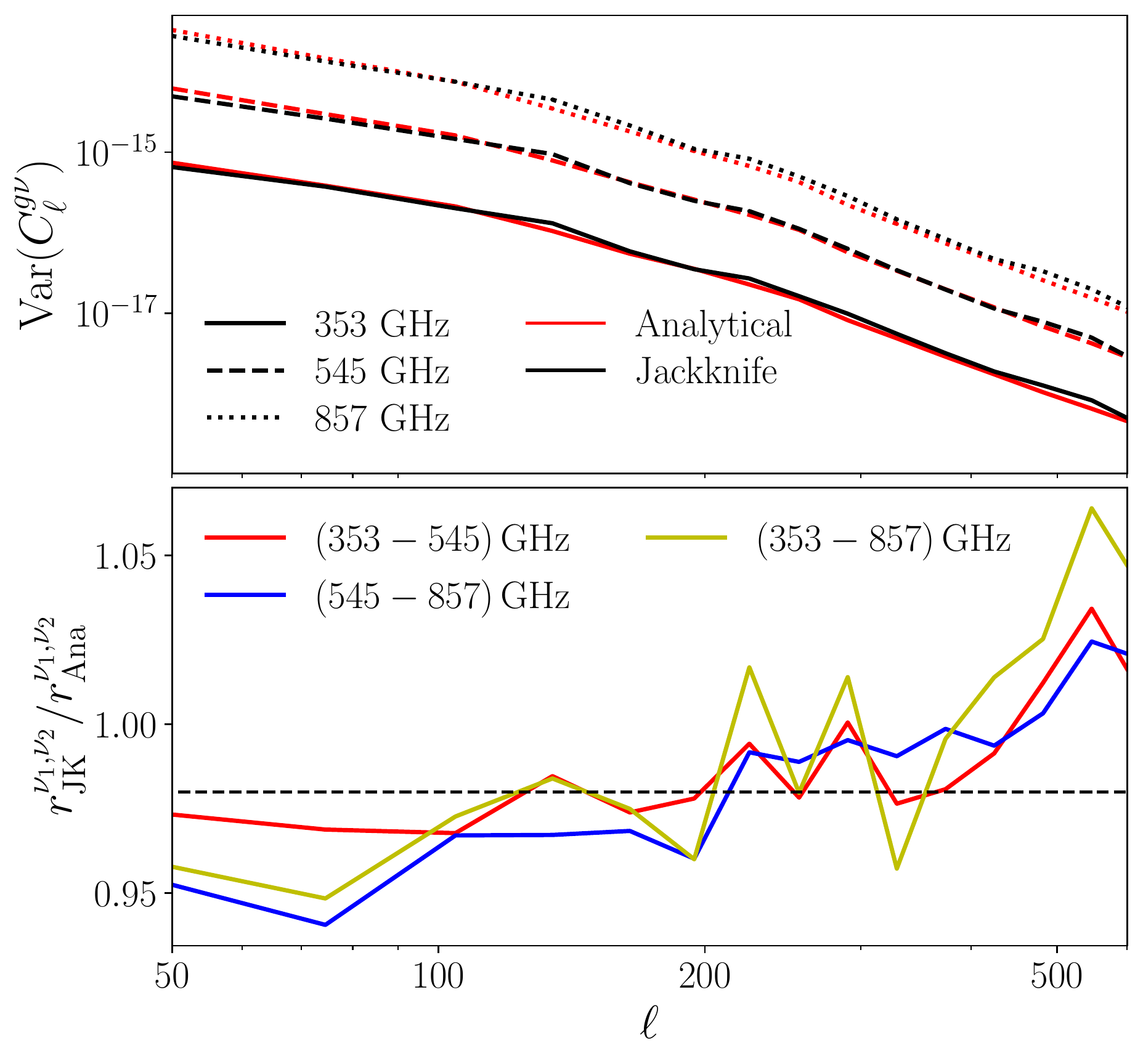}
      \caption{{\sl Top}: variance of the galaxy-CIB cross-correlation for the fourth \dls redshift bin with the three CIB frequencies (solid, dashed, and dotted lines respectively). Results are shown in red for the analytical covariance, and in black for the jackknife estimate. {\sl Bottom}: ratio between jackknife and analytical estimates for the correlation coefficient of pairs of cross-power spectra involving different CIB frequencies. Results are shown for the $(353-545)$ GHz, $(545-857)$ GHz and $(353-857)$ GHz pairs in red, blue and yellow respectively. We find that, on average, the jackknife estimate predicts a $\sim2\%$ lower correlation coefficient between frequencies. We correct our analytical estimate for this factor (shown as a horizontal dashed line).}\label{fig:cov_jk}
    \end{figure} %
    We measure the galaxy auto-correlation power spectrum for the six redshift bins considered, as well as their cross-correlations with the three CIB frequency maps, for a total of 24 auto- and cross-spectra. As an example, the galaxy clustering auto-correlations and the cross-correlations with the $545$ GHz CIB map are shown in Fig. \ref{fig:fitscuts}. The measured power spectra are shown in black together with their statistical uncertainties. The light grey bands show the fiducial scale cuts used in our analysis. We impose a default low-$\ell$ cut of $\ell_{\rm min}=100$. This is mostly motivated by the loss of large-scale power in the CIB maps caused by the removal of Galactic contamination on different large-area patches, as reported by \citet{Lenz19}. Although this should not affect the galaxy auto-correlation, observational systematics leading to artificial contamination of the galaxy number counts across the sky may lead to extra power on the largest scales. $\ell_{\rm min}=100$ is likely a conservative choice in this case \citep{2010.00466,2021JCAP...10..030G}, but we choose to keep the same cut for both power spectra for simplicity. On small scales, we remove all multipoles above $\ell_{\rm max}=k_{\rm max}\bar{\chi}-1/2$, where $\bar{\chi}$ is the radial comoving distance to the mean redshift of each galaxy sample, and $k_{\rm max}=0.15\,{\rm Mpc}^{-1}$. As we argued in Section \ref{sssec:methods.theo.tomo}, we expect the phenomenological model of Eq. \ref{eq:cl_pheno} to provide a good fit to the data on these scales. To test the robustness of our results to this choice, we will also present results for a less conservative cut of $k_{\rm max}=0.2\,{\rm Mpc}^{-1}$ (shown by the dark grey bands in the Figure). After these scale cuts, the complete data vector, comprising six galaxy auto-correlations and 6$\times$3 galaxy-CIB cross-correlations contains $200$ elements.
    
    Figure~\ref{fig:fitscuts} also shows, in red, the best-fit theoretical prediction following the model of Eq. \ref{eq:cl_pheno}. The residuals between data and prediction normalised by the diagonal errors are shown at the bottom of each panel. As discussed and quantified in the next section, overall we find that the model is able to provide a good fit to the data.

    Figure \ref{fig:covmat} shows the correlation matrix, defined in terms of the covariance $C_{ij}$ as $r_{ij}\equiv C_{ij}/\sqrt{C_{ii}C_{jj}}$. The blocks along the diagonal correspond to the six different redshift bins, with the galaxy auto-correlation followed by the 3 CIB cross-correlations. In each block, the cross-covariance between $C_\ell^{gg}$ and the $C_\ell^{g\nu}$s is significantly smaller than the cross-covariance between CIB channels. The latter is at the level of $\sim95\%$. This is because the three \planck{} maps trace essentially the same underlying CIB fluctuations with little frequency decorrelation. This makes an accurate calculation of the full covariance crucial since this high level of correlation requires the model residuals between different frequencies to follow the same trend almost exactly. The overlaps between the \dls redshift distributions shown in Figure \ref{fig:zbins} also lead to an appreciable cross-covariance between the power spectra in adjacent redshift bins.

    We validate the analytical estimate of the covariance matrix described in Section \ref{ssec:methods.map2cl} through jackknife resampling, dividing the footprint into 413 jackknife regions. In general, we find good agreement between both estimates of the covariance matrix, and we are able to recover the diagonal errors within $\sim4-5\%$ (see top panel of Fig. \ref{fig:cov_jk}). However, we find that the jackknife estimate recovers cross-correlation coefficients between power spectra involving different CIB frequencies that are consistently lower than the analytical prediction by $\sim2\%$ (see bottom panel of Fig. \ref{fig:cov_jk}). Although this is a small difference, the tight correlation between different CIB frequencies makes it necessary to accurately characterise these off-diagonal covariance elements. We find that, otherwise, the goodness of fit to the full dataset combining all frequencies was unacceptable, even if the same model parameters were able to provide a good fit to the three different frequencies separately, which clearly points to an inconsistent covariance matrix. To correct for this, we therefore multiply all off-diagonal covariance matrix elements involving two different CIB frequencies by an overall factor of 0.98. The difference is likely due to inhomogeneity in the map noise across the footprint, which is not accounted for in the analytical covariance.

  \subsection{Model-independent measurements of $\bsfr$}\label{ssec:results.bsfr}
    \begin{figure*}
        \centering
        \includegraphics[width=0.99\textwidth]{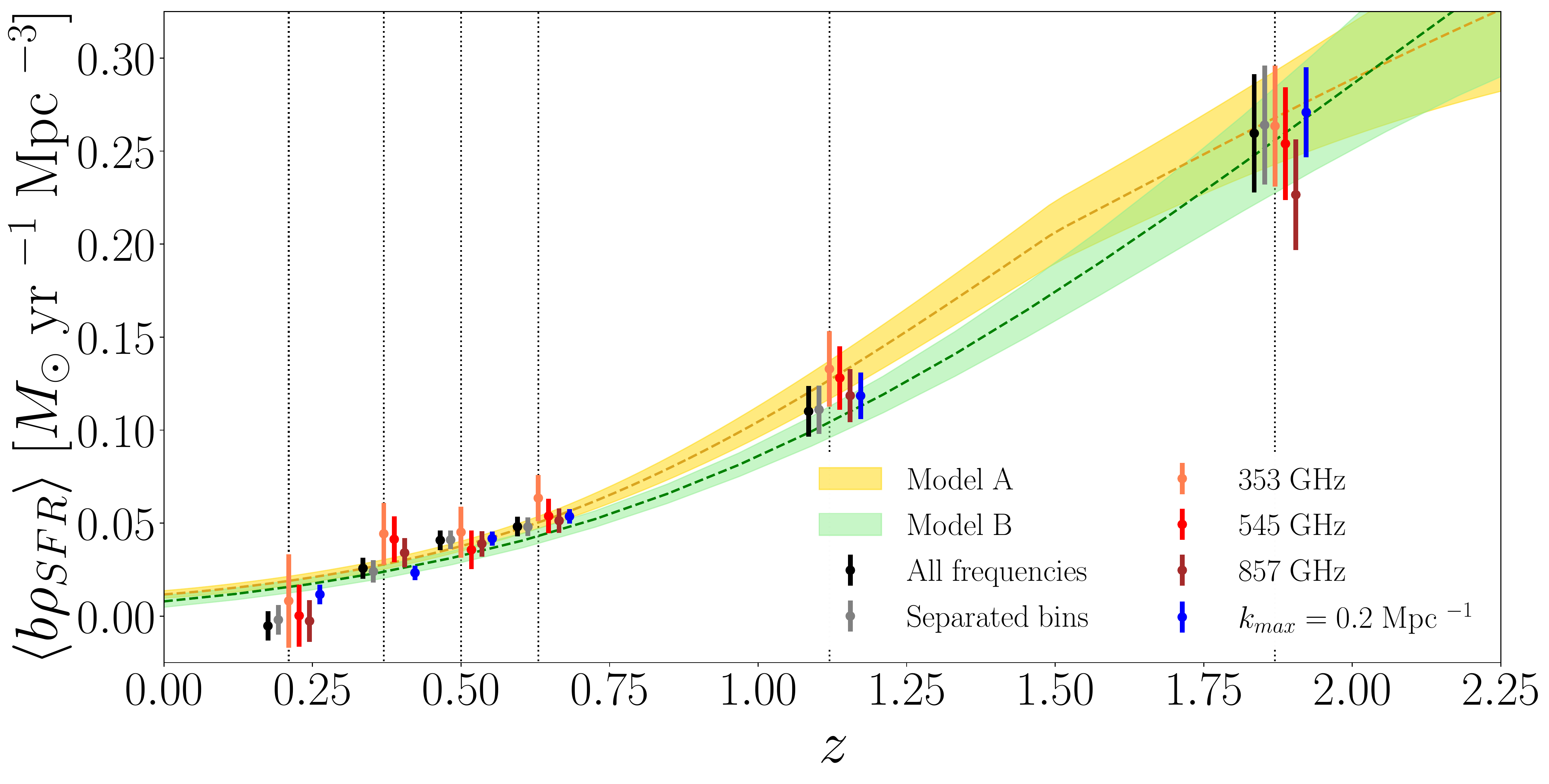}
        \caption{Inferred values of $\bsfr$ as a function of redshift for different analysis choices, with their $1\sigma$ error bars. Our fiducial measurements are shown in black. The results assuming less restrictive scale cuts are shown in blue, and the measurements carried out in each redshift bin independently are shown in grey. The constraints found from different frequency maps independently are shown in orange, red, and burgundy (for 353, 545, and 857 GHz respectively). The yellow and green bands show the prediction and 1-$\sigma$ confidence interval for $\bsfr$ obtained from the two halo models described in Section \ref{sssec:methods.theo.cib_sfr} (Model A and Model B respectively).}
        \label{fig:blin}
    \end{figure*} %
    \begin{figure*}
        \centering
        \includegraphics[width=0.7\textwidth]{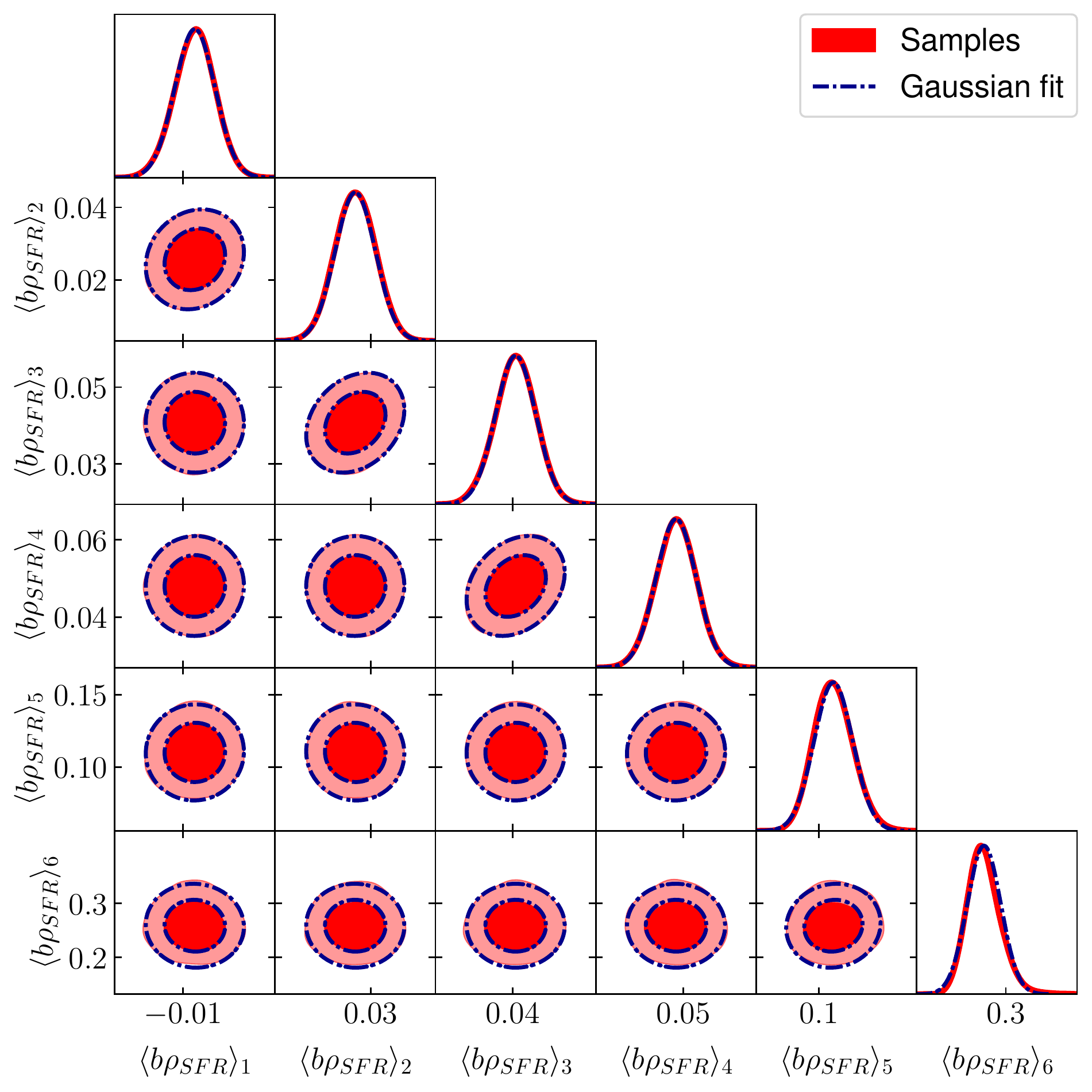}
        \caption{Posterior probability distributions for our measurements of $\bsfr$ in each redshift bin. The 1- and 2-$\sigma$ contours calculated from the MCMC chains are shown in red. The blue dash-dotted line shows the same contours for a multivariate normal distribution with the same mean and covariance as the chains. This justifies the use of a Gaussian likelihood when using these measurements to place constraints on halo-based models of the SFR in Section \ref{ssec:results.const_hm}.}
        \label{fig:bias}
    \end{figure*} %
    \begin{table}
        \centering
        \def\arraystretch{1.2}
        \begin{tabular}{|ccccc|}
            \hline
            & & $k_{\rm max}=0.15\,{\rm Mpc}^{-1}$ & & \\
            \hline
            \multirow{2}{*}{Bin} & \multirow{2}{*}{$\bar{z}$} & \multirow{2}{*}{$b_g$} & $\bsfr$ \\
            & & & ($\usfr$)\\
            \hline
            \multicolumn{5}{c|}{\textbf{\dls}} \\
            1   & 0.21 & $1.068\pm 0.044$ & $-0.005\pm 0.008$ \\
            2   & 0.37 & $1.382\pm 0.026$ & $0.026\pm 0.006$ \\
            3   & 0.50 & $1.306\pm 0.017$ & $0.041\pm 0.005$ \\
            4   & 0.63 & $1.733\pm 0.019$ & $0.048\pm 0.005$ \\
            \multicolumn{5}{c|}{\textbf{eBOSS}} \\
            5   & 1.12 & $2.057\pm 0.149$ & $0.110\pm 0.014$ \\ 
            6   & 1.87 & $2.165\pm 0.191$ & $0.260\pm 0.032$ \\
            \hline
            \multicolumn{4}{c|}{$\chi^{2}/N_{\rm d.o.f}=201.2/188 = 1.07, \hspace{12pt}{\rm PTE}=0.245$}\\
            \hline
        \end{tabular}
        \caption{Summary of the obtained values for the galaxy bias and the bias weighted star-formation rate within each redshift bin as presented in Figure \ref{fig:zbins}, when all parameters are sampled jointly. We display the mean values and the $1 \sigma$ uncertainties. The last row show the $\chi^{2}$ per degree of freedom value and PTE for our constraints.
        }\label{tab:joint}
    \end{table} %
    We obtain tomographic measurements of the bias-weighted mean SFR density, $\bsfr$, by fitting the phenomenological model in Eq. \ref{eq:cl_pheno} to the power spectrum data described in the previous section. The model consists on one galaxy bias parameter $b_g$ and one value of $\bsfr$ for each redshift bin, as well as a free white-noise amplitude ($n_{xy}$) for each individual power spectrum. Although the white noise levels for different CIB frequencies should be tightly correlated, we choose to allow each of them to vary freely, in order to account for different levels of point-source subtraction, and non-Poissonian stochastic contributions in galaxy clustering. The full model therefore has a total of $6\times N_{\rm bin}=36$ free parameters. Note, however, that since the white noise amplitudes are linear parameters, it is possible to marginalise over them analytically, significantly reducing the number of degrees of freedom we have to explore to $2\times N_{\rm bin}=12$. This is achieved by modifying the inverse covariance matrix used in the Gaussian likelihood of Eq. \ref{eq:like} as \citep{1992ApJ...398..169R}
    \begin{equation}
      \left({\sf C}_{\bf d}^{-1}\right)'={\sf C}_{\bf d}^{-1}-{\sf C}_{\bf d}^{-1}{\sf T}\left({\sf T}^T{\sf C}_{\bf d}^{-1}{\sf T}\right)^{-1}{\sf T}^T{\sf C}_{\bf d}^{-1},
    \end{equation}
    where ${\sf T}$ is a matrix containing, as columns, the templates corresponding to the different linear parameters of the model (in this case, a constant vector for each of the different power spectra included in the data vector). We used flat, uninformative priors on all parameters.

    Figure \ref{fig:blin} shows the resulting measurements of $\bsfr$. Our fiducial measurements, simultaneously using all redshift bins and CIB frequencies, with a small-scale cut of $k_{\rm max}=0.15\,{\rm Mpc}^{-1}$ are shown in black. The associated numerical values can be found in Table \ref{tab:joint}. We obtain $\sim10\%$ constraints on $\bsfr$ at $z\gtrsim0.5$, and a $\sim20\%$ measurement in the second redshift bin ($z\simeq0.37$). The conservative scale cuts used here leave only 3 data points in the cross-correlation with the first redshift bin, and we are only able to provide an upper bound on $\bsfr$, due to the degeneracy between this parameter and the white noise amplitude. The model is able to reproduce the data with a good $\chi^2$, with a probability-to-exceed ${\rm PTE}=0.245$. The full posterior distribution of the $\bsfr$ parameters is shown in Fig. \ref{fig:bias}. Although the measurements are largely statistically independent from one another, we can appreciate a non-negligible correlation between the measurements in adjacent \dls bins (e.g. bins 2-3 and 3-4). More importantly, we find that the posterior distribution is highly Gaussian. The dot-dashed black lines in the Figure show a multivariate Gaussian distribution with the same mean and covariance as that obtained from the MCMC chains, which reproduces the true posterior almost exactly. This will enable us to use the measured values of $\bsfr$ to constrain halo-based models of the star formation history in Section \ref{ssec:results.const_hm}. For convenience, we provide the correlation matrix of these measurements:
    \begin{equation}
      {\sf r}=\left(
      \begin{array}{cccccc}
        1.00  & 0.14  & -0.01 & 0.01  & 0.02  & -0.03 \\
        0.14  & 1.00  & 0.26  & 0.02  & -0.02 & -0.01 \\
        -0.01 & 0.26  & 1.00  & 0.27  & 0.01  & 0.00  \\
        0.01  & 0.02  & 0.27  & 1.00  & 0.02  & -0.01 \\
        0.02  & -0.02 & 0.01  & 0.02  & 1.00  & 0.06  \\
        -0.03 & -0.01 & 0.00  & -0.01 & 0.06  & 1.00
      \end{array}
      \right).
    \end{equation}

    To validate these measurements, we repeat our analysis changing the number of redshift bins and CIB frequencies used, as well as the small-scale cut. The results for each of these alternative analyses are listed in Tables \ref{tab:binsep}, \ref{tab:freqsep}, and \ref{tab:k=02} of Appendix \ref{app:tables}. First, in order to quantify the effect of the overlap between different redshift bins, we obtain new measurements of $\bsfr$ for each bin separately. We recover constraints that are almost equal to our fiducial measurements, and the model is able to reproduce the data for each redshift bin individually, with $\chi^2$ $p$-values above $0.1$. These results are shown as grey points in Fig. \ref{fig:blin}.
    
    The model used here relies on having an accurate description of the mean infrared spectral energy distribution ($S^{\rm eff}_\nu$ in Eq. \ref{eq:kernel_cib}), and inaccuracies in it would lead to inconsistent measurements of $\bsfr$ for different frequencies. To test for this, we repeat our measurement of $\bsfr$ three more times, using only one of the three CIB frequencies at a time.  The results are presented in Table \ref{tab:freqsep}, and as orange/red/brown points in Fig. \ref{fig:blin} for the 353/545/857 GHz channels respectively. Overall we find good agreement between the three frequencies. The 353 GHz channel achieves consistently larger uncertainties than the other two, and the recovered values of $\bsfr$ are also consistently higher, although compatible with the fiducial measurements within $1\sigma$. Although this could be a hint for a small inconsistency between the CIB data and the effective spectra of \citet{2013A&A...557A..66B,2015A&A...573A.113B,2017A&A...607A..89B}, it may also be an indication of contamination by other extragalactic foregrounds at lower frequencies, particularly the thermal Sunyaev-Zel'dovich effect. Since we do not observe a definite trend in the difference between the 353 GHz constraints and the rest as a function of redshift, and since the differences lie within the 68\% uncertainties, we did not investigate this further, and proceeded to combine all frequencies in our fiducial analysis.

    Finally we repeat the analysis relaxing the small-scale cut to $k_{\rm max}=0.2\,{\rm Mpc}^{-1}$, which allows us to include one or two additional bandpowers in each redshift bin (see Fig. \ref{fig:fitscuts}). Overall, the recovered values of $\bsfr$ are again in good agreement with our fiducial results, within $1\sigma$ uncertainties. With these additional data points, the uncertainties on $\bsfr$ reduce by $\sim20-30\%$. More importantly, the extra data point for the first \dls redshift bins allows us to break the degeneracy with the white noise amplitude and obtain a measurement of $\bsfr$ at $z\simeq0.2$. The value recovered is in reasonable agreement with the predictions of the best-fit halo models derived in the next Section when extrapolated to lower redshifts. The model is still able to provide a reasonable fit to the data with this less conservative scale cut (${\rm PTE}=4\%$). Our measurements of $\bsfr$ are therefore robust to the choice of scale cut made for our analysis.

  \subsection{Constraints on halo-based SFR models}\label{ssec:results.const_hm}
    \begin{figure*}
        \centering
        \includegraphics[width=0.48\textwidth]{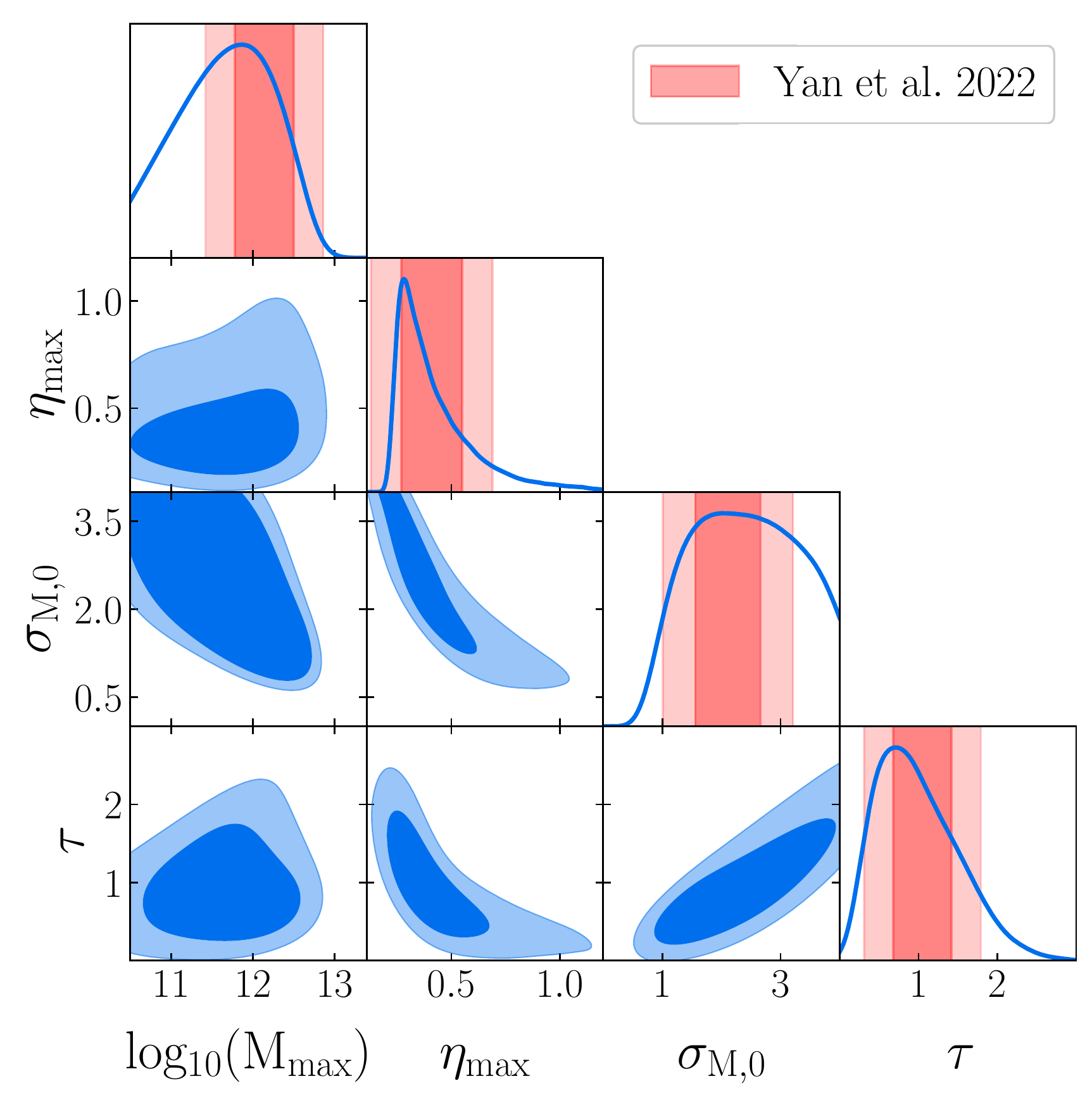}
        \includegraphics[width=0.48\textwidth]{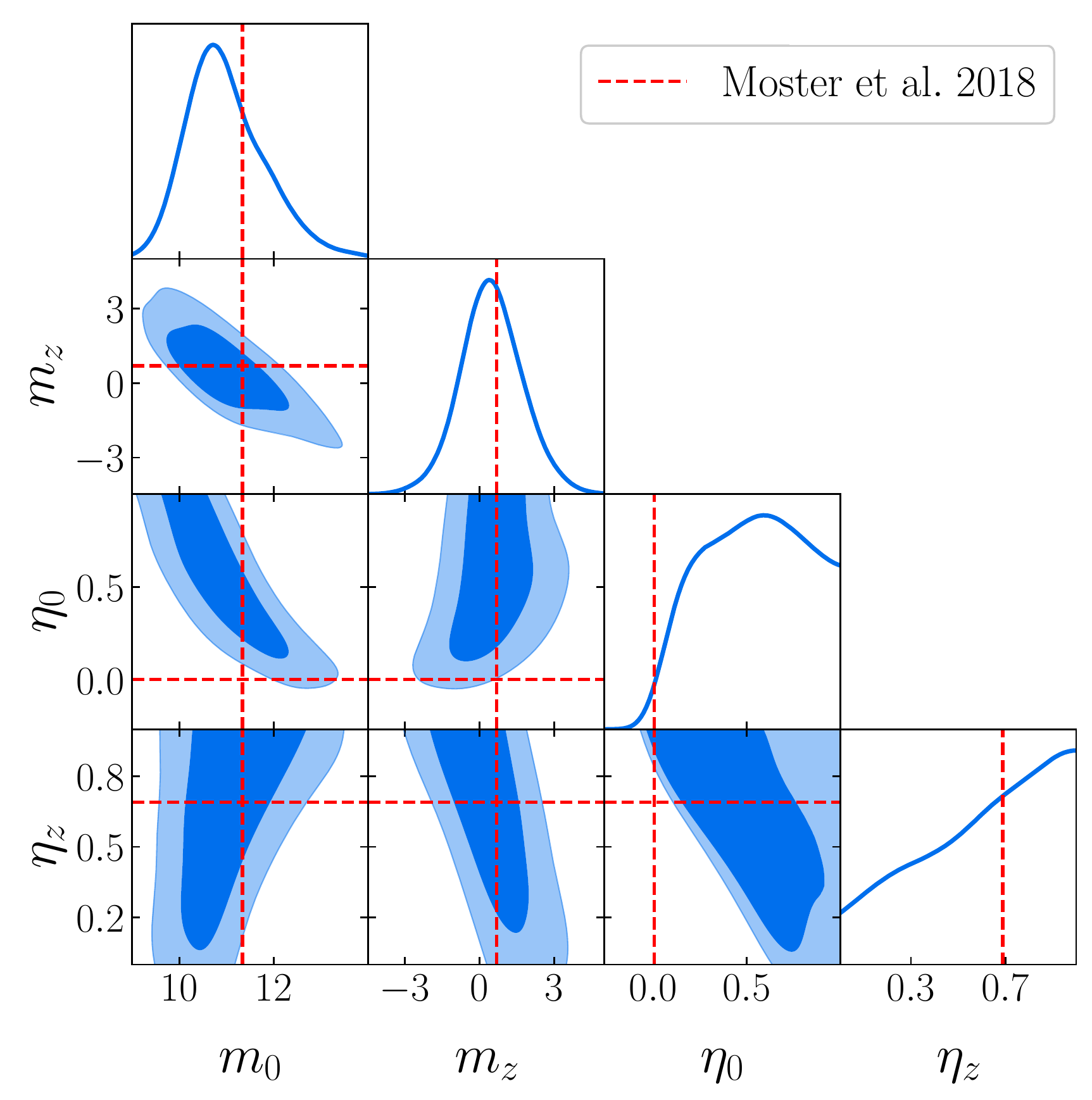}
        \caption{Constraints on the two halo-based SFR models (Model A and Model B in the left and right panels, respectively). The red bands in the left panel show the $1\sigma$ and $2\sigma$ constraints on Model A obtained by \citet{2022arXiv220401649Y}. The red lines in the right panel show the best-fit parameters obtained by \citet{2018MNRAS.477.1822M} for Model B.}
        \label{fig:constraints}
    \end{figure*}
    \begin{figure*}
        \centering
        \includegraphics[width=0.99\textwidth]{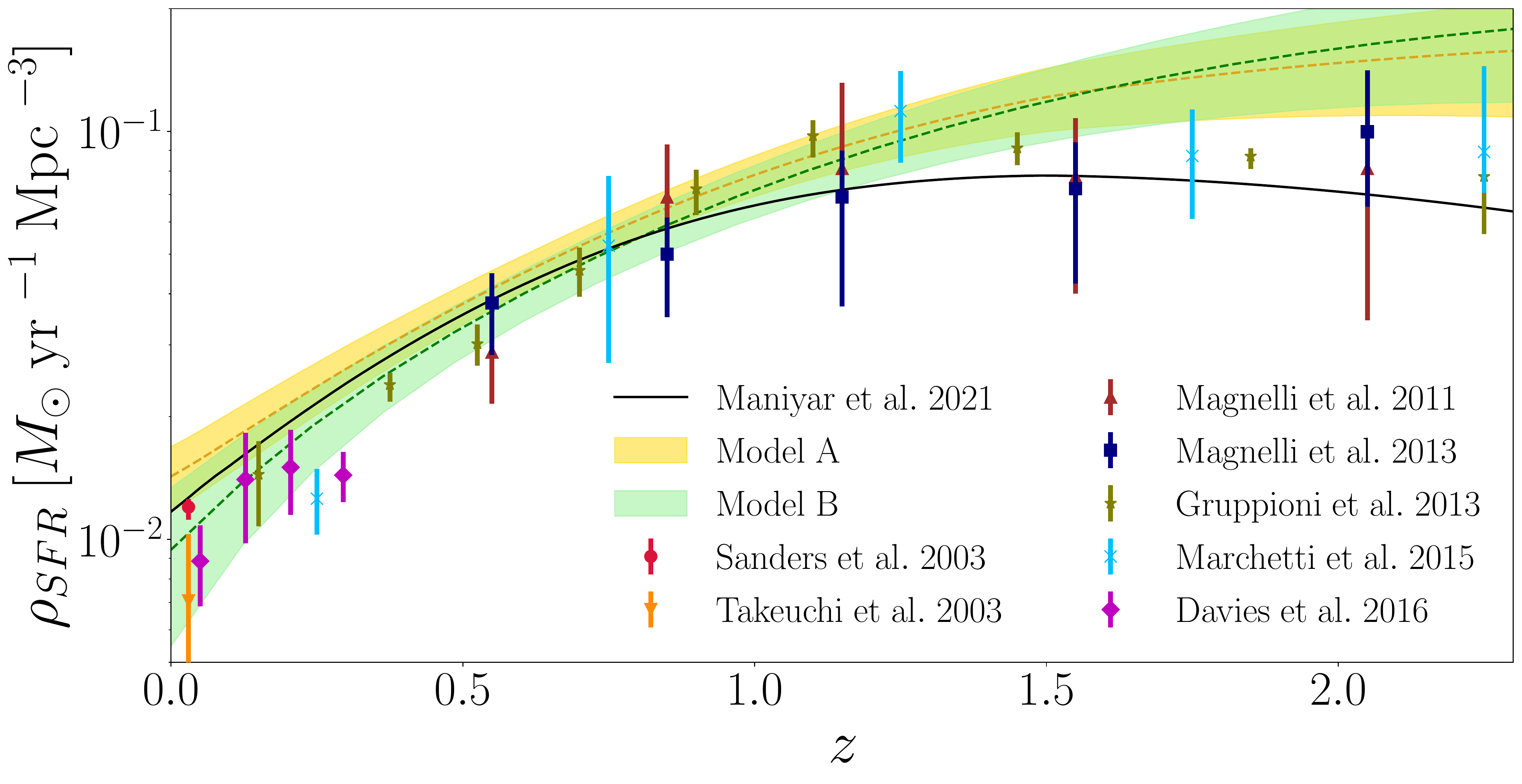}
        \caption{Star-formation rate density as a function of redshift. The points with error bars correspond to the direct measurements from the infrared luminosity function, extracted from \citet{madau2014cosmic, davies2016gama}. The solid black shows the best-fit prediction from the CIB analysis of \citet{2021A&A...645A..40M}. The two bands show the 1-$\sigma$ constraints obtained from our measurements of $\bsfr$ using the two halo-based SFR models described in Section \ref{sssec:methods.theo.cib_sfr} (Model A in yellow, Model B in green).}
        \label{fig:SFRD}
    \end{figure*}
    We use the tomographic measurements of $\bsfr$ presented in the previous section to derive constraints on the two halo models for the SFR history described in Section \ref{sssec:methods.theo.cib_sfr}. For this, we use again the Gaussian likelihood of Eq. \ref{eq:like}, where ${\bf d}$ is the set of six measured values of $\bsfr$, ${\sf C}_{\bf d}$ is the covariance of these measurements estimated from the MCMC chains, and ${\bf m}$ is the theoretical model, given by Eq. \ref{eq:bsfr}. The model has 4 free parameters for both models:
    \begin{itemize}
      \item {\bf Model A} has $\Theta=\{\eta_{\rm max},\log_{10}M_{\rm max},\sigma_{M,0},\tau\}$ with the following top-hat priors:
      \begin{align}
        &0 \leq \eta_{\rm max} \leq 1.2,\\
        &10.5 \leq \log_{10}M_{\rm max} \leq 14, \\
        &0 \leq \sigma_{M,0} \leq 4, \\
        &0 \leq \tau \leq 3.
      \end{align}
      \item {\bf Model B} has $\Theta=\{m_0,m_z,\eta_0,\eta_z\}$ with the following top-hat priors:
      \begin{align}
        & 9 \leq m_0 \leq 14,\\
        &-6 \leq m_z \leq 6, \\
        & 0 \leq \eta_0 \leq 1, \\
        &0 \leq \eta_z \leq 1.
      \end{align}
    \end{itemize}
    Since aiming to constrain 4 parameters with only 6 data points will likely lead to overfitting, we do not expect to be able to obtain competitive constraints on all of these parameters. Our main aim, however, is to use these models to map our measurements of $\bsfr$ onto constraints on the evolution of the star-formation rate density $\rsfr(z)$.

    The posterior distribution of the parameters of Model A is shown in Fig. \ref{fig:constraints} (left panel). Our constraints are in good agreement with those found by \citet{2022arXiv220401649Y} using the same CIB maps and galaxy clustering data from KiDS. These are shown as red vertical bands in the diagonal panels of the figure. Our measurements seem to favour slightly lower values of $\tau$ than those of \citet{2022arXiv220401649Y}, albeit with a long tail towards large $\tau$ that makes them broadly compatible. The peak efficiency $\eta_{\rm max}\sim0.4$ is found to be in good agreement with both \citet{2022arXiv220401649Y} and the CIB constraints from \citet{2021A&A...645A..40M}. On the other hand, our data favour a peak star formation mass of $\log_{10}M_{\rm max}\sim11.5$, significantly lower than that found by \citet{2021A&A...645A..40M} ($\log_{10}M_{\rm max}=12.94\pm0.02$), but compatible with the constraints of \citet{2022arXiv220401649Y}.

    The posterior distribution for Model B is shown in Fig. \ref{fig:constraints} (right panel). Our constraints are in good agreement with the best-fit parameters of \citet{2018MNRAS.477.1822M}\footnote{The constraints found by \citet{2018MNRAS.477.1822M} using a large dataset are much tighter than those found here, and thus we only show the final best-fit values.}, shown as dashed red lines. The data are able to constrain the peak mass parameters to
    \begin{equation}
      m_0=11.50\pm0.85,\hspace{12pt}m_z=0.5\pm1.3,
    \end{equation}
    although the peak efficiency parameters are largely unconstrained within their prior due to the degeneracy with $m_0$ and $m_z$. The peak mass is roughly constant, with $\log_{10}M_{\rm max}\simeq 11.6\pm0.8$ within the redshift range of our data. This is consistent with the results found with Model A.

    In order to place constraints on the star-formation rate history from our measurements of $\bsfr$, we randomly select 1000 samples from the MCMC chains for both models and calculate $\bsfr$ and $\rsfr$ as a function of redshift via Eqs. \ref{eq:bsfr} and \ref{eq:rhosfr} respectively. The yellow and green bands in Fig. \ref{fig:blin} show the $68\%$ confidence interval for the predicted evolution of $\bsfr$ extracted from Model A and Model B respectively. The corresponding constraints on the star-formation rate density are shown in Fig. \ref{fig:SFRD}. The figure also shows various direct measurements of $\rsfr$ from the infrared luminosity function \citep{2012MNRAS.421.2832S} (points with error bars), as well as the prediction from the best-fit model found by \citet{2021A&A...645A..40M} from the CIB auto-correlation. The direct $\rsfr$ measurements are from \citet{Sanders_2003,Takeuchi_2003,magnelli2011,2013A&A...553A.132M,2013MNRAS.432...23G,2016MNRAS.456.1999M} and \citet{2016MNRAS.461..458D}, and were extracted from \citet{madau2014cosmic} and \citet{davies2016gama}. Overall, our constraints are in rough agreement with these independent direct measurements with two main caveats. At low redshifts, our constraints on Model A predict a marginally higher SFR density compared with the direct measurements, whereas the predictions of Model B agree well with the data. This may be due to the fact that, unlike Model A, Model B allows for both the peak efficiency and the maximum efficiency mass to depend on redshift. The conversion from $\bsfr$ to $\rsfr$ at low-redshift prediction is therefore somewhat sensitive to the details of the underlying model. On the other hand, at high redshift both models predict a marginally higher $\rsfr$ compared with the direct measurements. This is likely driven by the large amplitude of the cross-correlation with the high-redshift eBOSS quasar sample, and therefore cross-correlations with other high-redshift tracers would be valuable to confirm this result.
    
    These caveats aside, the fact that our measurements of $\bsfr$ are compatible with direct measurements of the SFR density is an interesting and non-trivial result, beyond the fact that both constraints come from independent datasets. The contribution of haloes of different masses to $\bsfr$ is weighted by their linear bias, and thus $\bsfr$ is more sensitive than $\rsfr$ to star formation in the most massive haloes. An agreement between measurements of both quantities within a given model thus provides non-trivial support for the physical interpretation of the dependence of star formation on halo mass. The measurements of $\bsfr$ presented here, and those that will be achievable with future datasets, can therefore be used to strengthen constraints on star formation and galaxy evolution models, two areas of astrophysics of critical importance for cosmology in the next decade.

\section{Conclusions}\label{sec:conc}
  We have presented a tomographic analysis of Cosmic Infrared Background maps through cross-correlation with a suite of six galaxy samples spanning the redshift range $z\lesssim2$ comprising galaxies from the DESI Legacy Survey and quasars from eBOSS. The CIB is a good tracer of the obscured star formation history, and thus the main motivation behind this exercise was to place direct constraints on the evolution of star formation in the last $\sim10$ billion years. Our analysis has been carried out in two steps.
  
  First, we used a phenomenological parametrisation of the angular power spectra, inspired on the halo model, which allows us to directly measure the mean star-formation rate density weighted by halo mass, $\bsfr$, in a virtually model-independent manner (see Eq. \ref{eq:cl_pheno}). Although this involves restricting our analysis to the largest scales, thus discarding significant amounts of information, we have argued that the resulting constraints are robust to a wide range of theoretical uncertainties that a full modelling of the small-scale $C_\ell$s would be sensitive to. These include accurately describing the scale dependence of the mean and variance of the galaxy overdensity in haloes, the covariance between galaxy overdensity and SFR density in haloes (or, equivalently, characterising star formation in the target galaxy sample), the transition between the 1-halo and 2-halo-dominated regimes, and the different contributions (Poisson or otherwise) to the shot-noise component of all auto- and cross-correlations involved.
  
  We find that this phenomenological model provides a good fit to the data on the scales probed. We thus obtained measurements for $\bsfr$ at six different redshifts, reported in Table \ref{tab:joint}. The measurements are robust with respect to the choice of scale cuts, as well as the combination of datasets used (i.e. the choice of redshift bins and frequency maps). Furthermore, we have shown that the posterior distribution for these $\bsfr$ measurements is well approximated by a multivariate normal. It is therefore easy to use these measurements to derive constraints on any halo-based parametrisation of star formation using a simple Gaussian likelihood. Since $\bsfr$ is more sensitive to the contribution to the SFR density from massive haloes than direct measurements of $\rsfr$, a combination of both measurements can be used to put constraints on the halo mass dependence of star formation.

  As a second step, we use the measured values of $\bsfr$ as data to constrain the parameters of two halo-based models for the SFR history, characterising the mass and redshift dependence of the efficiency with which gas is converted into stars (see Fig. \ref{fig:eta_models}). Besides allowing us to place constraints on the models themselves, this procedure makes it possible to map our constraints on $\bsfr$ onto constraints on $\rsfr$ itself. The result is shown in Fig. \ref{fig:SFRD}. Our constraints are in good qualitative agreement with independent measurements of the SFR density. This supports the physical interpretation of the halo-based SFR models of \citet{2018MNRAS.477.1822M,2021A&A...645A..40M}, and the role of the CIB as a cosmic tracer of star formation. We have shown, however, that the predicted SFR density at low redshifts depends on the details of the underlying model, whereas our data seem to consistently predict a slightly higher $\rsfr$ at $z\sim2$. Thus our measurements of $\bsfr$ contain information that is complementary to that of direct measurements of the SFR density, and which can be used to improve our understanding of star formation in haloes.

  A number of caveats associated with our analysis must be noted. First and foremost, as Fig. \ref{fig:fitscuts} shows, the cross-correlation between the CIB and our galaxy samples is detected at very high significance over all scales and redshifts explored. Nevertheless, the simplified model used here requires us to discard most of these data. A more sophisticated model would allow us to derive significantly tighter constraints on both models of the star formation history, and on the astrophysical properties of the galaxy samples used. As we have noted above, several major modelling challenges must be overcome in order to construct such a model given the high statistical significant of the measurement. Ensuring that the model is able to recover unbiased constraints on its parameters would likely require the use of simulations incorporating a somewhat realistic description of star formation and galaxy evolution. The rewards of developing more sophisticated models describing this and other cross-correlations \citep{2015A&A...576A..90H,2017arXiv171110774H,2021JCAP...12..028K,2021PhRvD.103j3515M,2022JCAP...04..033K,2022A&A...660A..27T} are nevertheless worth the effort. For now, we have limited ourselves to providing these conservative but robust constraints, and leave a more complex analysis, modelling the small-scale power spectra directly for future work.

  Our model also relies on a number of simplified assumptions. First, we have assumed perfect knowledge of the mean infrared spectral energy distributions as tabulated by \citet{2013A&A...557A..66B,2015A&A...573A.113B,2017A&A...607A..89B}. Since our measurement is based on the large-scale bias extracted from the 2-halo term, it is insensitive to the scatter in this spectrum. However, any mis-modelling of the redshift dependence of the mean SED would directly lead to a biased recovery of the evolution of $\bsfr$ (although the fact that we recover compatible results from the three \planck{} frequencies is reassuring). Furthermore, we have assumed perfect knowledge of the relationship between infrared luminosity and SFR, through a simple proportionality constant (Eq. \ref{eq:luminosity.sfr}) that assumes particular universal initial mass function \citep{2003PASP..115..763C}. This assumption would affect our results and the direct measurements of $\rsfr$ in the same manner. The model also assumes that the SFR at a given time is proportional to the baryon accretion rate at the same time, whereas the dependence is likely more complicated. Our results are likely insensitive to most of these assumptions within the uncertainties reported, but they may become relevant in the presence of more data.

  In terms of systematics, the presence of other sources of extragalactic contamination in the CIB maps, which would correlate with the galaxy distribution, could bias the measured values of $\bsfr$. The most harmful contaminants in this case are thermal Sunyaev-Zel'dovich and emission from radio point sources. The fact that we recover compatible results from the three \planck{} frequencies reassures that any residual contamination is likely negligible within the range of scales and frequencies explored here. This may not be the case if a more ambitious analysis was carried out, using smaller-scale data or other frequencies, and a more thorough quantification of the impact from possible sources of contamination would have been necessary. On the galaxy side, Galactic extinction may cause a non-cosmological modulation in the observed number counts of galaxies. Since Galactic dust is also a contaminant of the CIB maps, the cross-correlation between both tracers may be affected by this systematic. In our case, the effect should be negligible given the conservative large-scale cut we have imposed, but a more careful study of this cross-contamination may be required for more sensitive data, or if the large-scale correlation of both tracers were to be exploited (as would be the case if targeting ultra-large scale cosmological science cases such as primordial non-Gaussianity \cite{2016MNRAS.463.2046T}). Finally, photometric redshift uncertainties leading to an incorrect determination of the galaxy redshift distribution may also impact the recovered constraints on $\bsfr$ \citep{2010.00466}. In our case this is only a potential concern for the \dls sample. Uncertainties in the sample's $p(z)$ can arise qualitatively in two ways: shifts in the mean of the distribution and changes in its width. Given the broad kernel of the CIB, our constraints should be impervious to moderate mean shifts. An incorrect determination of the width, however, would affect the amplitude of the galaxy auto-correlation, leaving the cross-correlation with the CIB almost untouched \citep{2021MNRAS.502..876A}. This would lead to an incorrect determination of the galaxy bias, which would then translate into a biased $\bsfr$. However, by examining the cross-bin correlations of the \dls sample used here, \citet{2010.00466} were able to tightly constrain the overlap between bins (which would change with the $p(z)$ width), and therefore this effect should be small in our analysis. A careful calibration of the redshift distribution would be needed, however, when attempting this measurement with other photometric samples.

  Further progress in the study of the CIB, and in its use as a cosmological tool, will become possible with the advent of next-generation experiments. On the CIB side, high-resolution data from experiments such as CCAT-prime \citep{2018SPIE10700E..1MS} or the Simons Observatory \citep{2019JCAP...02..056A} will improve the quality, sensitivity, and frequency coverage of current CIB observations. On the galaxy side, future deep photometric surveys such as the Rubin Observatory Legacy Survey of Space and Time \citep[LSST,][]{2009arXiv0912.0201L} will enable us to trace the large-scale structure tomographically to redshifts $z\sim3$, significantly improving our ability to trace the evolution of star formation at earlier times. Likewise, spectroscopic surveys such as DESI \citep{2019BAAS...51g..57L} will provide denser quasar samples, enabling the same type of constraints with better control over redshift uncertainties. With a better understanding of the star formation history, it will also be possible to use the CIB itself as a cosmological tool, allowing us to map the Universe's structures on very large scales with a sensitivity higher than will be achievable by future CMB lensing experiments. However, a large number of modelling and data analysis challenges need to be tackled before we can fully exploit the rich information encoded in the CIB.

\section*{Acknowledgements}
  We would like to thank Jens Chluba, Martin Rey, Aprajita Verma, and Ziang Yan for useful discussions. BJ is supported by the ENS Paris-Saclay. CGG is supported by European Research Council Grant No:  693024 and the Beecroft Trust. DA is supported by the Science and Technology Facilities Council through an Ernest Rutherford Fellowship, grant reference ST/P004474. JRZ and NK are supported by an STFC doctoral studentship. NK also acknowledges funding from the University of Oxford, Department of Physics. We also made extensive use of computational resources at the University of Oxford Department of Physics, funded by the John Fell Oxford University Press Research Fund. 

  We made extensive use of the {\tt numpy} \citep{oliphant2006guide, van2011numpy}, {\tt scipy} \citep{2020SciPy-NMeth}, {\tt astropy} \citep{1307.6212, 1801.02634}, {\tt healpy} \citep{Zonca2019}, {\tt GetDist} \cite{2019arXiv191013970L}, and {\tt matplotlib} \citep{Hunter:2007} python packages.

  The Legacy Surveys consist of three individual and complementary projects: the Dark Energy Camera Legacy Survey (DECaLS; Proposal ID \#2014B-0404; PIs: David Schlegel and Arjun Dey), the Beijing-Arizona Sky Survey (BASS; NOAO Prop. ID \#2015A-0801; PIs: Zhou Xu and Xiaohui Fan), and the Mayall z-band Legacy Survey (MzLS; Prop. ID \#2016A-0453; PI: Arjun Dey). DECaLS, BASS and MzLS together include data obtained, respectively, at the Blanco telescope, Cerro Tololo Inter-American Observatory, NSF's NOIRLab; the Bok telescope, Steward Observatory, University of Arizona; and the Mayall telescope, Kitt Peak National Observatory, NOIRLab. The Legacy Surveys project is honored to be permitted to conduct astronomical research on Iolkam Du\'ag (Kitt Peak), a mountain with particular significance to the Tohono O’odham Nation.

  NOIRLab is operated by the Association of Universities for Research in Astronomy (AURA) under a cooperative agreement with the National Science Foundation.

  This project used data obtained with the Dark Energy Camera (DECam), which was constructed by the Dark Energy Survey (DES) collaboration. Funding for the DES Projects has been provided by the U.S. Department of Energy, the U.S. National Science Foundation, the Ministry of Science and Education of Spain, the Science and Technology Facilities Council of the United Kingdom, the Higher Education Funding Council for England, the National Center for Supercomputing Applications at the University of Illinois at Urbana-Champaign, the Kavli Institute of Cosmological Physics at the University of Chicago, Center for Cosmology and Astro-Particle Physics at the Ohio State University, the Mitchell Institute for Fundamental Physics and Astronomy at Texas A\&M University, Financiadora de Estudos e Projetos, Fundacao Carlos Chagas Filho de Amparo, Financiadora de Estudos e Projetos, Fundacao Carlos Chagas Filho de Amparo a Pesquisa do Estado do Rio de Janeiro, Conselho Nacional de Desenvolvimento Cientifico e Tecnologico and the Ministerio da Ciencia, Tecnologia e Inovacao, the Deutsche Forschungsgemeinschaft and the Collaborating Institutions in the Dark Energy Survey. The Collaborating Institutions are Argonne National Laboratory, the University of California at Santa Cruz, the University of Cambridge, Centro de Investigaciones Energeticas, Medioambientales y Tecnologicas-Madrid, the University of Chicago, University College London, the DES-Brazil Consortium, the University of Edinburgh, the Eidgenossische Technische Hochschule (ETH) Zurich, Fermi National Accelerator Laboratory, the University of Illinois at Urbana-Champaign, the Institut de Ciencies de l'Espai (IEEC/CSIC), the Institut de Fisica d'Altes Energies, Lawrence Berkeley National Laboratory, the Ludwig Maximilians Universitat Munchen and the associated Excellence Cluster Universe, the University of Michigan, NSF's NOIRLab, the University of Nottingham, the Ohio State University, the University of Pennsylvania, the University of Portsmouth, SLAC National Accelerator Laboratory, Stanford University, the University of Sussex, and Texas A\&M University.

  BASS is a key project of the Telescope Access Program (TAP), which has been funded by the National Astronomical Observatories of China, the Chinese Academy of Sciences (the Strategic Priority Research Program “The Emergence of Cosmological Structures” Grant \# XDB09000000), and the Special Fund for Astronomy from the Ministry of Finance. The BASS is also supported by the External Cooperation Program of Chinese Academy of Sciences (Grant \# 114A11KYSB20160057), and Chinese National Natural Science Foundation (Grant \# 11433005).

  The Legacy Survey team makes use of data products from the Near-Earth Object Wide-field Infrared Survey Explorer (NEOWISE), which is a project of the Jet Propulsion Laboratory/California Institute of Technology. NEOWISE is funded by the National Aeronautics and Space Administration.

  The Legacy Surveys imaging of the DESI footprint is supported by the Director, Office of Science, Office of High Energy Physics of the U.S. Department of Energy under Contract No. DE-AC02-05CH1123, by the National Energy Research Scientific Computing Center, a DOE Office of Science User Facility under the same contract; and by the U.S. National Science Foundation, Division of Astronomical Sciences under Contract No. AST-0950945 to NOAO.

  Funding for the Sloan Digital Sky Survey IV has been provided by the Alfred P. Sloan Foundation, the U.S. Department of Energy Office of Science, and the Participating Institutions. 

  SDSS-IV acknowledges support and resources from the Center for High Performance Computing  at the University of Utah. The SDSS website is www.sdss.org.

  SDSS-IV is managed by the Astrophysical Research Consortium for the Participating Institutions of the SDSS Collaboration including the Brazilian Participation Group, the Carnegie Institution for Science, Carnegie Mellon University, Center for Astrophysics | Harvard \& Smithsonian, the Chilean Participation Group, the French Participation Group, Instituto de Astrof\'isica de Canarias, The Johns Hopkins University, Kavli Institute for the Physics and Mathematics of the Universe (IPMU) / University of Tokyo, the Korean Participation Group, Lawrence Berkeley National Laboratory, Leibniz Institut f\"ur Astrophysik Potsdam (AIP),  Max-Planck-Institut f\"ur Astronomie (MPIA Heidelberg), Max-Planck-Institut f\"ur Astrophysik (MPA Garching), Max-Planck-Institut f\"ur Extraterrestrische Physik (MPE), National Astronomical Observatories of China, New Mexico State University, New York University, University of Notre Dame, Observat\'ario Nacional / MCTI, The Ohio State University, Pennsylvania State University, Shanghai Astronomical Observatory, United Kingdom Participation Group, Universidad Nacional Aut\'onoma de M\'exico, University of Arizona, University of Colorado Boulder, University of Oxford, University of Portsmouth, University of Utah, University of Virginia, University of Washington, University of Wisconsin, Vanderbilt University, and Yale University. 

\section*{Data Availability}
  The code developed for this work as well as the derived data sets produced (power spectra and covariances) are available upon request. The catalogues and maps used were made publicly available by the authors, as described in the text.

\bibliographystyle{mnras}
\bibliography{main}

\begin{thebibliography}{}
\makeatletter
\relax
\def\mn@urlcharsother{\let\do\@makeother \do\$\do\&\do\#\do\^\do\_\do\%\do\~}
\def\mn@doi{\begingroup\mn@urlcharsother \@ifnextchar [ {\mn@doi@}
  {\mn@doi@[]}}
\def\mn@doi@[#1]#2{\def\@tempa{#1}\ifx\@tempa\@empty \href
  {http://dx.doi.org/#2} {doi:#2}\else \href {http://dx.doi.org/#2} {#1}\fi
  \endgroup}
\def\mn@eprint#1#2{\mn@eprint@#1:#2::\@nil}
\def\mn@eprint@arXiv#1{\href {http://arxiv.org/abs/#1} {{\tt arXiv:#1}}}
\def\mn@eprint@dblp#1{\href {http://dblp.uni-trier.de/rec/bibtex/#1.xml}
  {dblp:#1}}
\def\mn@eprint@#1:#2:#3:#4\@nil{\def\@tempa {#1}\def\@tempb {#2}\def\@tempc
  {#3}\ifx \@tempc \@empty \let \@tempc \@tempb \let \@tempb \@tempa \fi \ifx
  \@tempb \@empty \def\@tempb {arXiv}\fi \@ifundefined
  {mn@eprint@\@tempb}{\@tempb:\@tempc}{\expandafter \expandafter \csname
  mn@eprint@\@tempb\endcsname \expandafter{\@tempc}}}

\bibitem[\protect\citeauthoryear{{Ade} et~al.,}{{Ade}
  et~al.}{2019}]{2019JCAP...02..056A}
{Ade} P.,  et~al., 2019, \mn@doi [\jcap] {10.1088/1475-7516/2019/02/056}, \href
  {https://ui.adsabs.harvard.edu/abs/2019JCAP...02..056A} {2019, 056}

\bibitem[\protect\citeauthoryear{{Alonso}, {Sanchez}, {Slosar}  \& {LSST Dark
  Energy Science Collaboration}}{{Alonso} et~al.}{2019}]{1809.09603}
{Alonso} D.,  {Sanchez} J.,  {Slosar} A.,   {LSST Dark Energy Science
  Collaboration} 2019, \mn@doi [\mnras] {10.1093/mnras/stz093}, \href
  {https://ui.adsabs.harvard.edu/abs/2019MNRAS.484.4127A} {484, 4127}

\bibitem[\protect\citeauthoryear{{Alonso}, {Bellini}, {Hale}, {Jarvis}  \&
  {Schwarz}}{{Alonso} et~al.}{2021}]{2021MNRAS.502..876A}
{Alonso} D.,  {Bellini} E.,  {Hale} C.,  {Jarvis} M.~J.,   {Schwarz} D.~J.,
  2021, \mn@doi [\mnras] {10.1093/mnras/stab046}, \href
  {https://ui.adsabs.harvard.edu/abs/2021MNRAS.502..876A} {502, 876}

\bibitem[\protect\citeauthoryear{{Astropy Collaboration} et~al.,}{{Astropy
  Collaboration} et~al.}{2013}]{1307.6212}
{Astropy Collaboration} et~al., 2013, \mn@doi [\aap]
  {10.1051/0004-6361/201322068}, \href
  {https://ui.adsabs.harvard.edu/abs/2013A&A...558A..33A} {558, A33}

\bibitem[\protect\citeauthoryear{{Astropy Collaboration} et~al.,}{{Astropy
  Collaboration} et~al.}{2018}]{1801.02634}
{Astropy Collaboration} et~al., 2018, \mn@doi [\aj] {10.3847/1538-3881/aabc4f},
  \href {https://ui.adsabs.harvard.edu/abs/2018AJ....156..123A} {156, 123}

\bibitem[\protect\citeauthoryear{Bekhti et~al.,}{Bekhti
  et~al.}{2016}]{bekhti2016hi4pi}
Bekhti N.~B.,  et~al., 2016, Astronomy \& Astrophysics, 594, A116

\bibitem[\protect\citeauthoryear{Berlind \& Weinberg}{Berlind \&
  Weinberg}{2002}]{berlind2002halo}
Berlind A.~A.,  Weinberg D.~H.,  2002, The Astrophysical Journal, 575, 587

\bibitem[\protect\citeauthoryear{{B{\'e}thermin}, {Wang}, {Dor{\'e}},
  {Lagache}, {Sargent}, {Daddi}, {Cousin}  \& {Aussel}}{{B{\'e}thermin}
  et~al.}{2013}]{2013A&A...557A..66B}
{B{\'e}thermin} M.,  {Wang} L.,  {Dor{\'e}} O.,  {Lagache} G.,  {Sargent} M.,
  {Daddi} E.,  {Cousin} M.,   {Aussel} H.,  2013, \mn@doi [\aap]
  {10.1051/0004-6361/201321688}, \href
  {https://ui.adsabs.harvard.edu/abs/2013A&A...557A..66B} {557, A66}

\bibitem[\protect\citeauthoryear{{B{\'e}thermin} et~al.,}{{B{\'e}thermin}
  et~al.}{2015}]{2015A&A...573A.113B}
{B{\'e}thermin} M.,  et~al., 2015, \mn@doi [\aap]
  {10.1051/0004-6361/201425031}, \href
  {https://ui.adsabs.harvard.edu/abs/2015A&A...573A.113B} {573, A113}

\bibitem[\protect\citeauthoryear{{B{\'e}thermin} et~al.,}{{B{\'e}thermin}
  et~al.}{2017}]{2017A&A...607A..89B}
{B{\'e}thermin} M.,  et~al., 2017, \mn@doi [\aap]
  {10.1051/0004-6361/201730866}, \href
  {https://ui.adsabs.harvard.edu/abs/2017A&A...607A..89B} {607, A89}

\bibitem[\protect\citeauthoryear{{Blum} et~al.,}{{Blum}
  et~al.}{2016}]{2016AAS...22831701B}
{Blum} R.~D.,  et~al., 2016, in American Astronomical Society Meeting Abstracts
  \#228. p. 317.01

\bibitem[\protect\citeauthoryear{{Cao}, {Gong}, {Feng}, {Cooray}, {Cheng}  \&
  {Chen}}{{Cao} et~al.}{2020}]{2020ApJ...901...34C}
{Cao} Y.,  {Gong} Y.,  {Feng} C.,  {Cooray} A.,  {Cheng} G.,   {Chen} X.,
  2020, \mn@doi [\apj] {10.3847/1538-4357/abada1}, \href
  {https://ui.adsabs.harvard.edu/abs/2020ApJ...901...34C} {901, 34}

\bibitem[\protect\citeauthoryear{{Chabrier}}{{Chabrier}}{2003}]{2003PASP..115..763C}
{Chabrier} G.,  2003, \mn@doi [\pasp] {10.1086/376392}, \href
  {https://ui.adsabs.harvard.edu/abs/2003PASP..115..763C} {115, 763}

\bibitem[\protect\citeauthoryear{{Chen} et~al.,}{{Chen}
  et~al.}{2016}]{2016ApJ...831...91C}
{Chen} C.-C.,  et~al., 2016, \mn@doi [\apj] {10.3847/0004-637X/831/1/91}, \href
  {https://ui.adsabs.harvard.edu/abs/2016ApJ...831...91C} {831, 91}

\bibitem[\protect\citeauthoryear{{Chiang}, {M{\'e}nard}  \&
  {Schiminovich}}{{Chiang} et~al.}{2019}]{2019ApJ...877..150C}
{Chiang} Y.-K.,  {M{\'e}nard} B.,   {Schiminovich} D.,  2019, \mn@doi [\apj]
  {10.3847/1538-4357/ab1b35}, \href
  {https://ui.adsabs.harvard.edu/abs/2019ApJ...877..150C} {877, 150}

\bibitem[\protect\citeauthoryear{{Chiang}, {Makiya}, {M{\'e}nard}  \&
  {Komatsu}}{{Chiang} et~al.}{2020}]{2020ApJ...902...56C}
{Chiang} Y.-K.,  {Makiya} R.,  {M{\'e}nard} B.,   {Komatsu} E.,  2020, \mn@doi
  [\apj] {10.3847/1538-4357/abb403}, \href
  {https://ui.adsabs.harvard.edu/abs/2020ApJ...902...56C} {902, 56}

\bibitem[\protect\citeauthoryear{{Chisari} et~al.,}{{Chisari}
  et~al.}{2019}]{1812.05995}
{Chisari} N.~E.,  et~al., 2019, \mn@doi [\apjs] {10.3847/1538-4365/ab1658},
  \href {https://ui.adsabs.harvard.edu/abs/2019ApJS..242....2C} {242, 2}

\bibitem[\protect\citeauthoryear{Cooray \& Sheth}{Cooray \&
  Sheth}{2002}]{cooray2002halo}
Cooray A.,  Sheth R.,  2002, Physics reports, 372, 1

\bibitem[\protect\citeauthoryear{{Darwish}, {Sherwin}, {Sailer}, {Schaan}  \&
  {Ferraro}}{{Darwish} et~al.}{2021a}]{2021arXiv211100462D}
{Darwish} O.,  {Sherwin} B.~D.,  {Sailer} N.,  {Schaan} E.,   {Ferraro} S.,
  2021a, arXiv e-prints, \href
  {https://ui.adsabs.harvard.edu/abs/2021arXiv211100462D} {p. arXiv:2111.00462}

\bibitem[\protect\citeauthoryear{{Darwish} et~al.,}{{Darwish}
  et~al.}{2021b}]{2021MNRAS.500.2250D}
{Darwish} O.,  et~al., 2021b, \mn@doi [\mnras] {10.1093/mnras/staa3438}, \href
  {https://ui.adsabs.harvard.edu/abs/2021MNRAS.500.2250D} {500, 2250}

\bibitem[\protect\citeauthoryear{{Davies} et~al.,}{{Davies}
  et~al.}{2016a}]{2016MNRAS.461..458D}
{Davies} L.~J.~M.,  et~al., 2016a, \mn@doi [\mnras] {10.1093/mnras/stw1342},
  \href {https://ui.adsabs.harvard.edu/abs/2016MNRAS.461..458D} {461, 458}

\bibitem[\protect\citeauthoryear{Davies et~al.,}{Davies
  et~al.}{2016b}]{davies2016gama}
Davies L.~J.,  et~al., 2016b, Monthly Notices of the Royal Astronomical
  Society, 461, 458

\bibitem[\protect\citeauthoryear{{Dey} et~al.,}{{Dey}
  et~al.}{2019}]{1804.08657}
{Dey} A.,  et~al., 2019, \mn@doi [\aj] {10.3847/1538-3881/ab089d}, \href
  {https://ui.adsabs.harvard.edu/abs/2019AJ....157..168D} {157, 168}

\bibitem[\protect\citeauthoryear{{Dole} et~al.,}{{Dole}
  et~al.}{2006}]{2006A&A...451..417D}
{Dole} H.,  et~al., 2006, \mn@doi [\aap] {10.1051/0004-6361:20054446}, \href
  {https://ui.adsabs.harvard.edu/abs/2006A&A...451..417D} {451, 417}

\bibitem[\protect\citeauthoryear{{Dunkley} et~al.,}{{Dunkley}
  et~al.}{2013}]{2013JCAP...07..025D}
{Dunkley} J.,  et~al., 2013, \mn@doi [\jcap] {10.1088/1475-7516/2013/07/025},
  \href {https://ui.adsabs.harvard.edu/abs/2013JCAP...07..025D} {2013, 025}

\bibitem[\protect\citeauthoryear{{Dwek} et~al.,}{{Dwek}
  et~al.}{1998}]{1998ApJ...508..106D}
{Dwek} E.,  et~al., 1998, \mn@doi [\apj] {10.1086/306382}, \href
  {https://ui.adsabs.harvard.edu/abs/1998ApJ...508..106D} {508, 106}

\bibitem[\protect\citeauthoryear{{Fakhouri}, {Ma}  \&
  {Boylan-Kolchin}}{{Fakhouri} et~al.}{2010}]{2010MNRAS.406.2267F}
{Fakhouri} O.,  {Ma} C.-P.,   {Boylan-Kolchin} M.,  2010, \mn@doi [\mnras]
  {10.1111/j.1365-2966.2010.16859.x}, \href
  {https://ui.adsabs.harvard.edu/abs/2010MNRAS.406.2267F} {406, 2267}

\bibitem[\protect\citeauthoryear{{Fang}, {Eifler}, {Schaan}, {Huang}, {Krause}
  \& {Ferraro}}{{Fang} et~al.}{2022}]{2022MNRAS.509.5721F}
{Fang} X.,  {Eifler} T.,  {Schaan} E.,  {Huang} H.-J.,  {Krause} E.,
  {Ferraro} S.,  2022, \mn@doi [\mnras] {10.1093/mnras/stab3410}, \href
  {https://ui.adsabs.harvard.edu/abs/2022MNRAS.509.5721F} {509, 5721}

\bibitem[\protect\citeauthoryear{{Flaugher} et~al.,}{{Flaugher}
  et~al.}{2015}]{1504.02900}
{Flaugher} B.,  et~al., 2015, \mn@doi [\aj] {10.1088/0004-6256/150/5/150},
  \href {https://ui.adsabs.harvard.edu/abs/2015AJ....150..150F} {150, 150}

\bibitem[\protect\citeauthoryear{{Foreman-Mackey}, {Hogg}, {Lang}  \&
  {Goodman}}{{Foreman-Mackey} et~al.}{2013}]{emcee}
{Foreman-Mackey} D.,  {Hogg} D.~W.,  {Lang} D.,   {Goodman} J.,  2013, \mn@doi
  [PASP] {10.1086/670067}, 125, 306

\bibitem[\protect\citeauthoryear{{Garc{\'\i}a-Garc{\'\i}a}, {Alonso}  \&
  {Bellini}}{{Garc{\'\i}a-Garc{\'\i}a} et~al.}{2019}]{1906.11765}
{Garc{\'\i}a-Garc{\'\i}a} C.,  {Alonso} D.,   {Bellini} E.,  2019, \mn@doi
  [\jcap] {10.1088/1475-7516/2019/11/043}, \href
  {https://ui.adsabs.harvard.edu/abs/2019JCAP...11..043G} {2019, 043}

\bibitem[\protect\citeauthoryear{{Garc{\'\i}a-Garc{\'\i}a}, {Ruiz-Zapatero},
  {Alonso}, {Bellini}, {Ferreira}, {Mueller}, {Nicola}  \&
  {Ruiz-Lapuente}}{{Garc{\'\i}a-Garc{\'\i}a}
  et~al.}{2021}]{2021JCAP...10..030G}
{Garc{\'\i}a-Garc{\'\i}a} C.,  {Ruiz-Zapatero} J.,  {Alonso} D.,  {Bellini} E.,
   {Ferreira} P.~G.,  {Mueller} E.-M.,  {Nicola} A.,   {Ruiz-Lapuente} P.,
  2021, \mn@doi [\jcap] {10.1088/1475-7516/2021/10/030}, \href
  {https://ui.adsabs.harvard.edu/abs/2021JCAP...10..030G} {2021, 030}

\bibitem[\protect\citeauthoryear{{G{\'o}rski}, {Hivon}, {Banday}, {Wandelt},
  {Hansen}, {Reinecke}  \& {Bartelmann}}{{G{\'o}rski}
  et~al.}{2005}]{2005ApJ...622..759G}
{G{\'o}rski} K.~M.,  {Hivon} E.,  {Banday} A.~J.,  {Wandelt} B.~D.,  {Hansen}
  F.~K.,  {Reinecke} M.,   {Bartelmann} M.,  2005, \mn@doi [\apj]
  {10.1086/427976}, \href
  {https://ui.adsabs.harvard.edu/abs/2005ApJ...622..759G} {622, 759}

\bibitem[\protect\citeauthoryear{{Gruppioni} et~al.,}{{Gruppioni}
  et~al.}{2013}]{2013MNRAS.432...23G}
{Gruppioni} C.,  et~al., 2013, \mn@doi [\mnras] {10.1093/mnras/stt308}, \href
  {https://ui.adsabs.harvard.edu/abs/2013MNRAS.432...23G} {432, 23}

\bibitem[\protect\citeauthoryear{{Hamimeche} \& {Lewis}}{{Hamimeche} \&
  {Lewis}}{2008}]{2008PhRvD..77j3013H}
{Hamimeche} S.,  {Lewis} A.,  2008, \mn@doi [\prd]
  {10.1103/PhysRevD.77.103013}, \href
  {https://ui.adsabs.harvard.edu/abs/2008PhRvD..77j3013H} {77, 103013}

\bibitem[\protect\citeauthoryear{{Hang}, {Alam}, {Peacock}  \& {Cai}}{{Hang}
  et~al.}{2021}]{2010.00466}
{Hang} Q.,  {Alam} S.,  {Peacock} J.~A.,   {Cai} Y.-C.,  2021, \mn@doi [\mnras]
  {10.1093/mnras/staa3738}, \href
  {https://ui.adsabs.harvard.edu/abs/2021MNRAS.501.1481H} {501, 1481}

\bibitem[\protect\citeauthoryear{{Hivon}, {G{\'o}rski}, {Netterfield}, {Crill},
  {Prunet}  \& {Hansen}}{{Hivon} et~al.}{2002}]{astro-ph/0105302}
{Hivon} E.,  {G{\'o}rski} K.~M.,  {Netterfield} C.~B.,  {Crill} B.~P.,
  {Prunet} S.,   {Hansen} F.,  2002, \mn@doi [\apj] {10.1086/338126}, \href
  {https://ui.adsabs.harvard.edu/abs/2002ApJ...567....2H} {567, 2}

\bibitem[\protect\citeauthoryear{{Hou} et~al.,}{{Hou}
  et~al.}{2021}]{2007.08998}
{Hou} J.,  et~al., 2021, \mn@doi [\mnras] {10.1093/mnras/staa3234}, \href
  {https://ui.adsabs.harvard.edu/abs/2021MNRAS.500.1201H} {500, 1201}

\bibitem[\protect\citeauthoryear{Hunter}{Hunter}{2007}]{Hunter:2007}
Hunter J.~D.,  2007, \mn@doi [Computing in Science \& Engineering]
  {10.1109/MCSE.2007.55}, 9, 90

\bibitem[\protect\citeauthoryear{{Hurier}, {Douspis}, {Aghanim},
  {Pointecouteau}, {Diego}  \& {Macias-Perez}}{{Hurier}
  et~al.}{2015}]{2015A&A...576A..90H}
{Hurier} G.,  {Douspis} M.,  {Aghanim} N.,  {Pointecouteau} E.,  {Diego} J.~M.,
    {Macias-Perez} J.~F.,  2015, \mn@doi [\aap] {10.1051/0004-6361/201425555},
  \href {https://ui.adsabs.harvard.edu/abs/2015A&A...576A..90H} {576, A90}

\bibitem[\protect\citeauthoryear{{Hurier}, {Singh}  \&
  {Hern{\'a}ndez-Monteagudo}}{{Hurier} et~al.}{2017}]{2017arXiv171110774H}
{Hurier} G.,  {Singh} P.,   {Hern{\'a}ndez-Monteagudo} C.,  2017, arXiv
  e-prints, \href {https://ui.adsabs.harvard.edu/abs/2017arXiv171110774H} {p.
  arXiv:1711.10774}

\bibitem[\protect\citeauthoryear{{Kennicutt}}{{Kennicutt}}{1998}]{1998ARA&A..36..189K}
{Kennicutt} Robert~C. J.,  1998, \mn@doi [\araa]
  {10.1146/annurev.astro.36.1.189}, \href
  {https://ui.adsabs.harvard.edu/abs/1998ARA&A..36..189K} {36, 189}

\bibitem[\protect\citeauthoryear{{Kennicutt} \& {Evans}}{{Kennicutt} \&
  {Evans}}{2012}]{2012ARA&A..50..531K}
{Kennicutt} R.~C.,  {Evans} N.~J.,  2012, \mn@doi [\araa]
  {10.1146/annurev-astro-081811-125610}, \href
  {https://ui.adsabs.harvard.edu/abs/2012ARA&A..50..531K} {50, 531}

\bibitem[\protect\citeauthoryear{{Kere{\v{s}}}, {Katz}, {Weinberg}  \&
  {Dav{\'e}}}{{Kere{\v{s}}} et~al.}{2005}]{2005MNRAS.363....2K}
{Kere{\v{s}}} D.,  {Katz} N.,  {Weinberg} D.~H.,   {Dav{\'e}} R.,  2005,
  \mn@doi [\mnras] {10.1111/j.1365-2966.2005.09451.x}, \href
  {https://ui.adsabs.harvard.edu/abs/2005MNRAS.363....2K} {363, 2}

\bibitem[\protect\citeauthoryear{{Knox}, {Cooray}, {Eisenstein}  \&
  {Haiman}}{{Knox} et~al.}{2001}]{2001ApJ...550....7K}
{Knox} L.,  {Cooray} A.,  {Eisenstein} D.,   {Haiman} Z.,  2001, \mn@doi [\apj]
  {10.1086/319732}, \href
  {https://ui.adsabs.harvard.edu/abs/2001ApJ...550....7K} {550, 7}

\bibitem[\protect\citeauthoryear{{Koukoufilippas}, {Alonso}, {Bilicki}  \&
  {Peacock}}{{Koukoufilippas} et~al.}{2020}]{2020MNRAS.491.5464K}
{Koukoufilippas} N.,  {Alonso} D.,  {Bilicki} M.,   {Peacock} J.~A.,  2020,
  \mn@doi [\mnras] {10.1093/mnras/stz3351}, \href
  {https://ui.adsabs.harvard.edu/abs/2020MNRAS.491.5464K} {491, 5464}

\bibitem[\protect\citeauthoryear{{Krolewski} \& {Ferraro}}{{Krolewski} \&
  {Ferraro}}{2022}]{2022JCAP...04..033K}
{Krolewski} A.,  {Ferraro} S.,  2022, \mn@doi [\jcap]
  {10.1088/1475-7516/2022/04/033}, \href
  {https://ui.adsabs.harvard.edu/abs/2022JCAP...04..033K} {2022, 033}

\bibitem[\protect\citeauthoryear{{Krolewski}, {Ferraro}  \&
  {White}}{{Krolewski} et~al.}{2021}]{2021JCAP...12..028K}
{Krolewski} A.,  {Ferraro} S.,   {White} M.,  2021, \mn@doi [\jcap]
  {10.1088/1475-7516/2021/12/028}, \href
  {https://ui.adsabs.harvard.edu/abs/2021JCAP...12..028K} {2021, 028}

\bibitem[\protect\citeauthoryear{{LSST Science Collaboration} et~al.,}{{LSST
  Science Collaboration} et~al.}{2009}]{2009arXiv0912.0201L}
{LSST Science Collaboration} et~al., 2009, arXiv e-prints, \href
  {https://ui.adsabs.harvard.edu/abs/2009arXiv0912.0201L} {p. arXiv:0912.0201}

\bibitem[\protect\citeauthoryear{{Lenz}, {Dor{\'e}}  \& {Lagache}}{{Lenz}
  et~al.}{2019}]{Lenz19}
{Lenz} D.,  {Dor{\'e}} O.,   {Lagache} G.,  2019, \mn@doi [\apj]
  {10.3847/1538-4357/ab3c2b}, \href
  {https://ui.adsabs.harvard.edu/abs/2019ApJ...883...75L} {883, 75}

\bibitem[\protect\citeauthoryear{{Levi} et~al.,}{{Levi}
  et~al.}{2019}]{2019BAAS...51g..57L}
{Levi} M.,  et~al., 2019, in Bulletin of the American Astronomical Society.
  p.~57 (\mn@eprint {arXiv} {1907.10688})

\bibitem[\protect\citeauthoryear{{Lewis}}{{Lewis}}{2019}]{2019arXiv191013970L}
{Lewis} A.,  2019, arXiv e-prints, \href
  {https://ui.adsabs.harvard.edu/abs/2019arXiv191013970L} {p. arXiv:1910.13970}

\bibitem[\protect\citeauthoryear{{Limber}}{{Limber}}{1953}]{1953ApJ...117..134L}
{Limber} D.~N.,  1953, \mn@doi [\apj] {10.1086/145672}, \href
  {https://ui.adsabs.harvard.edu/abs/1953ApJ...117..134L} {117, 134}

\bibitem[\protect\citeauthoryear{Madau \& Dickinson}{Madau \&
  Dickinson}{2014a}]{madau2014cosmic}
Madau P.,  Dickinson M.,  2014a, arXiv preprint arXiv:1403.0007

\bibitem[\protect\citeauthoryear{{Madau} \& {Dickinson}}{{Madau} \&
  {Dickinson}}{2014b}]{2014ARA&A..52..415M}
{Madau} P.,  {Dickinson} M.,  2014b, \mn@doi [\araa]
  {10.1146/annurev-astro-081811-125615}, \href
  {https://ui.adsabs.harvard.edu/abs/2014ARA&A..52..415M} {52, 415}

\bibitem[\protect\citeauthoryear{{Magnelli}, {Elbaz, D.}, {Chary, R. R.},
  {Dickinson, M.}, {Le Borgne, D.}, {Frayer, D. T.}  \& {Willmer, C. N.
  A.}}{{Magnelli} et~al.}{2011}]{magnelli2011}
{Magnelli} {Elbaz, D.} {Chary, R. R.} {Dickinson, M.} {Le Borgne, D.} {Frayer,
  D. T.}  {Willmer, C. N. A.} 2011, \mn@doi [A\&A]
  {10.1051/0004-6361/200913941}, 528, A35

\bibitem[\protect\citeauthoryear{{Magnelli} et~al.,}{{Magnelli}
  et~al.}{2013}]{2013A&A...553A.132M}
{Magnelli} B.,  et~al., 2013, \mn@doi [\aap] {10.1051/0004-6361/201321371},
  \href {https://ui.adsabs.harvard.edu/abs/2013A&A...553A.132M} {553, A132}

\bibitem[\protect\citeauthoryear{Maniyar, Béthermin  \& Lagache}{Maniyar
  et~al.}{2018}]{Maniyar_2018}
Maniyar A.~S.,  Béthermin M.,   Lagache G.,  2018, \mn@doi [\aap]
  {10.1051/0004-6361/201732499}, 614, A39

\bibitem[\protect\citeauthoryear{{Maniyar}, {B{\'e}thermin}  \&
  {Lagache}}{{Maniyar} et~al.}{2021}]{2021A&A...645A..40M}
{Maniyar} A.,  {B{\'e}thermin} M.,   {Lagache} G.,  2021, \mn@doi [\aap]
  {10.1051/0004-6361/202038790}, \href
  {https://ui.adsabs.harvard.edu/abs/2021A&A...645A..40M} {645, A40}

\bibitem[\protect\citeauthoryear{{Marchetti} et~al.,}{{Marchetti}
  et~al.}{2016}]{2016MNRAS.456.1999M}
{Marchetti} L.,  et~al., 2016, \mn@doi [\mnras] {10.1093/mnras/stv2717}, \href
  {https://ui.adsabs.harvard.edu/abs/2016MNRAS.456.1999M} {456, 1999}

\bibitem[\protect\citeauthoryear{{McCarthy} \& {Madhavacheril}}{{McCarthy} \&
  {Madhavacheril}}{2021}]{2021PhRvD.103j3515M}
{McCarthy} F.,  {Madhavacheril} M.~S.,  2021, \mn@doi [\prd]
  {10.1103/PhysRevD.103.103515}, \href
  {https://ui.adsabs.harvard.edu/abs/2021PhRvD.103j3515M} {103, 103515}

\bibitem[\protect\citeauthoryear{{Mead} \& {Verde}}{{Mead} \&
  {Verde}}{2021}]{2021MNRAS.503.3095M}
{Mead} A.~J.,  {Verde} L.,  2021, \mn@doi [\mnras] {10.1093/mnras/stab748},
  \href {https://ui.adsabs.harvard.edu/abs/2021MNRAS.503.3095M} {503, 3095}

\bibitem[\protect\citeauthoryear{{Mead}, {Brieden}, {Tr{\"o}ster}  \&
  {Heymans}}{{Mead} et~al.}{2021}]{2021MNRAS.502.1401M}
{Mead} A.~J.,  {Brieden} S.,  {Tr{\"o}ster} T.,   {Heymans} C.,  2021, \mn@doi
  [\mnras] {10.1093/mnras/stab082}, \href
  {https://ui.adsabs.harvard.edu/abs/2021MNRAS.502.1401M} {502, 1401}

\bibitem[\protect\citeauthoryear{{M{\'e}nard}, {Scranton}, {Schmidt},
  {Morrison}, {Jeong}, {Budavari}  \& {Rahman}}{{M{\'e}nard}
  et~al.}{2013}]{2013arXiv1303.4722M}
{M{\'e}nard} B.,  {Scranton} R.,  {Schmidt} S.,  {Morrison} C.,  {Jeong} D.,
  {Budavari} T.,   {Rahman} M.,  2013, arXiv e-prints, \href
  {https://ui.adsabs.harvard.edu/abs/2013arXiv1303.4722M} {p. arXiv:1303.4722}

\bibitem[\protect\citeauthoryear{{Morrison}, {Hildebrandt}, {Schmidt},
  {Baldry}, {Bilicki}, {Choi}, {Erben}  \& {Schneider}}{{Morrison}
  et~al.}{2017}]{2017MNRAS.467.3576M}
{Morrison} C.~B.,  {Hildebrandt} H.,  {Schmidt} S.~J.,  {Baldry} I.~K.,
  {Bilicki} M.,  {Choi} A.,  {Erben} T.,   {Schneider} P.,  2017, \mn@doi
  [\mnras] {10.1093/mnras/stx342}, \href
  {https://ui.adsabs.harvard.edu/abs/2017MNRAS.467.3576M} {467, 3576}

\bibitem[\protect\citeauthoryear{{Moster}, {Naab}  \& {White}}{{Moster}
  et~al.}{2018}]{2018MNRAS.477.1822M}
{Moster} B.~P.,  {Naab} T.,   {White} S. D.~M.,  2018, \mn@doi [\mnras]
  {10.1093/mnras/sty655}, \href
  {https://ui.adsabs.harvard.edu/abs/2018MNRAS.477.1822M} {477, 1822}

\bibitem[\protect\citeauthoryear{{Neveux} et~al.,}{{Neveux}
  et~al.}{2020}]{2007.08999}
{Neveux} R.,  et~al., 2020, \mn@doi [\mnras] {10.1093/mnras/staa2780}, \href
  {https://ui.adsabs.harvard.edu/abs/2020MNRAS.499..210N} {499, 210}

\bibitem[\protect\citeauthoryear{{Newman}}{{Newman}}{2008}]{2008ApJ...684...88N}
{Newman} J.~A.,  2008, \mn@doi [\apj] {10.1086/589982}, \href
  {https://ui.adsabs.harvard.edu/abs/2008ApJ...684...88N} {684, 88}

\bibitem[\protect\citeauthoryear{{Nicola}, {Garc{\'\i}a-Garc{\'\i}a}, {Alonso},
  {Dunkley}, {Ferreira}, {Slosar}  \& {Spergel}}{{Nicola}
  et~al.}{2021}]{2010.09717}
{Nicola} A.,  {Garc{\'\i}a-Garc{\'\i}a} C.,  {Alonso} D.,  {Dunkley} J.,
  {Ferreira} P.~G.,  {Slosar} A.,   {Spergel} D.~N.,  2021, \mn@doi [\jcap]
  {10.1088/1475-7516/2021/03/067}, \href
  {https://ui.adsabs.harvard.edu/abs/2021JCAP...03..067N} {2021, 067}

\bibitem[\protect\citeauthoryear{Oliphant}{Oliphant}{2006}]{oliphant2006guide}
Oliphant T.~E.,  2006, A guide to NumPy.
~ Vol. 1, Trelgol Publishing USA

\bibitem[\protect\citeauthoryear{{Osborne}, {Hanson}  \& {Dor{\'e}}}{{Osborne}
  et~al.}{2014}]{2014JCAP...03..024O}
{Osborne} S.~J.,  {Hanson} D.,   {Dor{\'e}} O.,  2014, \mn@doi [\jcap]
  {10.1088/1475-7516/2014/03/024}, \href
  {https://ui.adsabs.harvard.edu/abs/2014JCAP...03..024O} {2014, 024}

\bibitem[\protect\citeauthoryear{{Pandey} et~al.,}{{Pandey}
  et~al.}{2019}]{2019PhRvD.100f3519P}
{Pandey} S.,  et~al., 2019, \mn@doi [\prd] {10.1103/PhysRevD.100.063519}, \href
  {https://ui.adsabs.harvard.edu/abs/2019PhRvD.100f3519P} {100, 063519}

\bibitem[\protect\citeauthoryear{{Partridge} \& {Peebles}}{{Partridge} \&
  {Peebles}}{1967}]{1967ApJ...147..868P}
{Partridge} R.~B.,  {Peebles} P.~J.~E.,  1967, \mn@doi [\apj] {10.1086/149079},
  \href {https://ui.adsabs.harvard.edu/abs/1967ApJ...147..868P} {147, 868}

\bibitem[\protect\citeauthoryear{Peacock \& Smith}{Peacock \&
  Smith}{2000}]{peacock2000halo}
Peacock J.,  Smith R.,  2000, Monthly Notices of the Royal Astronomical
  Society, 318, 1144

\bibitem[\protect\citeauthoryear{{Planck Collaboration} et~al.,}{{Planck
  Collaboration} et~al.}{2011}]{2011A&A...536A..18P}
{Planck Collaboration} et~al., 2011, \mn@doi [\aap]
  {10.1051/0004-6361/201116461}, \href
  {https://ui.adsabs.harvard.edu/abs/2011A&A...536A..18P} {536, A18}

\bibitem[\protect\citeauthoryear{{Planck Collaboration} et~al.,}{{Planck
  Collaboration} et~al.}{2014a}]{2014A&A...571A..18P}
{Planck Collaboration} et~al., 2014a, \mn@doi [\aap]
  {10.1051/0004-6361/201321540}, \href
  {https://ui.adsabs.harvard.edu/abs/2014A&A...571A..18P} {571, A18}

\bibitem[\protect\citeauthoryear{{Planck Collaboration} et~al.,}{{Planck
  Collaboration} et~al.}{2014b}]{2014A&A...571A..30P}
{Planck Collaboration} et~al., 2014b, \mn@doi [\aap]
  {10.1051/0004-6361/201322093}, \href
  {https://ui.adsabs.harvard.edu/abs/2014A&A...571A..30P} {571, A30}

\bibitem[\protect\citeauthoryear{{Planck Collaboration} et~al.,}{{Planck
  Collaboration} et~al.}{2016}]{2016A&A...596A.109P}
{Planck Collaboration} et~al., 2016, \mn@doi [\aap]
  {10.1051/0004-6361/201629022}, \href
  {https://ui.adsabs.harvard.edu/abs/2016A&A...596A.109P} {596, A109}

\bibitem[\protect\citeauthoryear{{Puget}, {Abergel}, {Bernard}, {Boulanger},
  {Burton}, {Desert}  \& {Hartmann}}{{Puget}
  et~al.}{1996}]{1996A&A...308L...5P}
{Puget} J.~L.,  {Abergel} A.,  {Bernard} J.~P.,  {Boulanger} F.,  {Burton}
  W.~B.,  {Desert} F.~X.,   {Hartmann} D.,  1996, \aap, \href
  {https://ui.adsabs.harvard.edu/abs/1996A&A...308L...5P} {308, L5}

\bibitem[\protect\citeauthoryear{{Ross} et~al.,}{{Ross}
  et~al.}{2020}]{2007.09000}
{Ross} A.~J.,  et~al., 2020, \mn@doi [\mnras] {10.1093/mnras/staa2416}, \href
  {https://ui.adsabs.harvard.edu/abs/2020MNRAS.498.2354R} {498, 2354}

\bibitem[\protect\citeauthoryear{{Rybicki} \& {Press}}{{Rybicki} \&
  {Press}}{1992}]{1992ApJ...398..169R}
{Rybicki} G.~B.,  {Press} W.~H.,  1992, \mn@doi [\apj] {10.1086/171845}, \href
  {https://ui.adsabs.harvard.edu/abs/1992ApJ...398..169R} {398, 169}

\bibitem[\protect\citeauthoryear{{Sailer}, {Schaan}, {Ferraro}, {Darwish}  \&
  {Sherwin}}{{Sailer} et~al.}{2021}]{2021PhRvD.104l3514S}
{Sailer} N.,  {Schaan} E.,  {Ferraro} S.,  {Darwish} O.,   {Sherwin} B.,  2021,
  \mn@doi [\prd] {10.1103/PhysRevD.104.123514}, \href
  {https://ui.adsabs.harvard.edu/abs/2021PhRvD.104l3514S} {104, 123514}

\bibitem[\protect\citeauthoryear{Sanders, Mazzarella, Kim, Surace  \&
  Soifer}{Sanders et~al.}{2003}]{Sanders_2003}
Sanders D.~B.,  Mazzarella J.~M.,  Kim D.-C.,  Surace J.~A.,   Soifer B.~T.,
  2003, \mn@doi [The Astronomical Journal] {10.1086/376841}, 126, 1607

\bibitem[\protect\citeauthoryear{{Schmittfull} \& {Seljak}}{{Schmittfull} \&
  {Seljak}}{2018}]{2018PhRvD..97l3540S}
{Schmittfull} M.,  {Seljak} U.,  2018, \mn@doi [\prd]
  {10.1103/PhysRevD.97.123540}, \href
  {https://ui.adsabs.harvard.edu/abs/2018PhRvD..97l3540S} {97, 123540}

\bibitem[\protect\citeauthoryear{{Scranton} et~al.,}{{Scranton}
  et~al.}{2005}]{2005ApJ...633..589S}
{Scranton} R.,  et~al., 2005, \mn@doi [\apj] {10.1086/431358}, \href
  {https://ui.adsabs.harvard.edu/abs/2005ApJ...633..589S} {633, 589}

\bibitem[\protect\citeauthoryear{Seljak}{Seljak}{2000}]{Seljak_2000}
Seljak U.,  2000, \mn@doi [Monthly Notices of the Royal Astronomical Society]
  {10.1046/j.1365-8711.2000.03715.x}, 318, 203

\bibitem[\protect\citeauthoryear{{Serra}, {Lagache}, {Dor{\'e}}, {Pullen}  \&
  {White}}{{Serra} et~al.}{2014}]{2014A&A...570A..98S}
{Serra} P.,  {Lagache} G.,  {Dor{\'e}} O.,  {Pullen} A.,   {White} M.,  2014,
  \mn@doi [\aap] {10.1051/0004-6361/201423958}, \href
  {https://ui.adsabs.harvard.edu/abs/2014A&A...570A..98S} {570, A98}

\bibitem[\protect\citeauthoryear{{Shang}, {Haiman}, {Knox}  \& {Oh}}{{Shang}
  et~al.}{2012}]{2012MNRAS.421.2832S}
{Shang} C.,  {Haiman} Z.,  {Knox} L.,   {Oh} S.~P.,  2012, \mn@doi [\mnras]
  {10.1111/j.1365-2966.2012.20510.x}, \href
  {https://ui.adsabs.harvard.edu/abs/2012MNRAS.421.2832S} {421, 2832}

\bibitem[\protect\citeauthoryear{{Silk}}{{Silk}}{2003}]{2003MNRAS.343..249S}
{Silk} J.,  2003, \mn@doi [\mnras] {10.1046/j.1365-8711.2003.06674.x}, \href
  {https://ui.adsabs.harvard.edu/abs/2003MNRAS.343..249S} {343, 249}

\bibitem[\protect\citeauthoryear{{Silva} et~al.,}{{Silva}
  et~al.}{2016}]{2016AAS...22831702S}
{Silva} D.~R.,  et~al., 2016, in American Astronomical Society Meeting
  Abstracts \#228. p. 317.02

\bibitem[\protect\citeauthoryear{{Stacey} et~al.,}{{Stacey}
  et~al.}{2018}]{2018SPIE10700E..1MS}
{Stacey} G.~J.,  et~al., 2018, in {Marshall} H.~K.,  {Spyromilio} J.,  eds,
  Society of Photo-Optical Instrumentation Engineers (SPIE) Conference Series
  Vol. 10700, Ground-based and Airborne Telescopes VII. p. 107001M (\mn@eprint
  {arXiv} {1807.04354}), \mn@doi{10.1117/12.2314031}

\bibitem[\protect\citeauthoryear{{Takada} \& {Hu}}{{Takada} \&
  {Hu}}{2013}]{2013PhRvD..87l3504T}
{Takada} M.,  {Hu} W.,  2013, \mn@doi [\prd] {10.1103/PhysRevD.87.123504},
  \href {https://ui.adsabs.harvard.edu/abs/2013PhRvD..87l3504T} {87, 123504}

\bibitem[\protect\citeauthoryear{{Takahashi}, {Sato}, {Nishimichi}, {Taruya}
  \& {Oguri}}{{Takahashi} et~al.}{2012}]{1208.2701}
{Takahashi} R.,  {Sato} M.,  {Nishimichi} T.,  {Taruya} A.,   {Oguri} M.,
  2012, \mn@doi [\apj] {10.1088/0004-637X/761/2/152}, \href
  {https://ui.adsabs.harvard.edu/abs/2012ApJ...761..152T} {761, 152}

\bibitem[\protect\citeauthoryear{Takeuchi, Yoshikawa  \& Ishii}{Takeuchi
  et~al.}{2003}]{Takeuchi_2003}
Takeuchi T.~T.,  Yoshikawa K.,   Ishii T.~T.,  2003, \mn@doi [The Astrophysical
  Journal] {10.1086/375181}, 587, L89

\bibitem[\protect\citeauthoryear{{Tinker} \& {Wetzel}}{{Tinker} \&
  {Wetzel}}{2010}]{2010ApJ...719...88T}
{Tinker} J.~L.,  {Wetzel} A.~R.,  2010, \mn@doi [\apj]
  {10.1088/0004-637X/719/1/88}, \href
  {https://ui.adsabs.harvard.edu/abs/2010ApJ...719...88T} {719, 88}

\bibitem[\protect\citeauthoryear{Tinker, Kravtsov, Klypin, Abazajian, Warren,
  Yepes, Gottlöber  \& Holz}{Tinker et~al.}{2008}]{Tinker_2008}
Tinker J.,  Kravtsov A.~V.,  Klypin A.,  Abazajian K.,  Warren M.,  Yepes G.,
  Gottlöber S.,   Holz D.~E.,  2008, \mn@doi [The Astrophysical Journal]
  {10.1086/591439}, 688, 709

\bibitem[\protect\citeauthoryear{{Tinker}, {Robertson}, {Kravtsov}, {Klypin},
  {Warren}, {Yepes}  \& {Gottl{\"o}ber}}{{Tinker}
  et~al.}{2010}]{2010ApJ...724..878T}
{Tinker} J.~L.,  {Robertson} B.~E.,  {Kravtsov} A.~V.,  {Klypin} A.,  {Warren}
  M.~S.,  {Yepes} G.,   {Gottl{\"o}ber} S.,  2010, \mn@doi [\apj]
  {10.1088/0004-637X/724/2/878}, \href
  {https://ui.adsabs.harvard.edu/abs/2010ApJ...724..878T} {724, 878}

\bibitem[\protect\citeauthoryear{{Tinsley}}{{Tinsley}}{1980}]{1980FCPh....5..287T}
{Tinsley} B.~M.,  1980, \mn@doi [\fcp] {10.48550/arXiv.2203.02041}, \href
  {https://ui.adsabs.harvard.edu/abs/1980FCPh....5..287T} {5, 287}

\bibitem[\protect\citeauthoryear{{Tr{\"o}ster} et~al.,}{{Tr{\"o}ster}
  et~al.}{2022}]{2022A&A...660A..27T}
{Tr{\"o}ster} T.,  et~al., 2022, \mn@doi [\aap] {10.1051/0004-6361/202142197},
  \href {https://ui.adsabs.harvard.edu/abs/2022A&A...660A..27T} {660, A27}

\bibitem[\protect\citeauthoryear{{Tucci}, {Desjacques}  \& {Kunz}}{{Tucci}
  et~al.}{2016}]{2016MNRAS.463.2046T}
{Tucci} M.,  {Desjacques} V.,   {Kunz} M.,  2016, \mn@doi [\mnras]
  {10.1093/mnras/stw2086}, \href
  {https://ui.adsabs.harvard.edu/abs/2016MNRAS.463.2046T} {463, 2046}

\bibitem[\protect\citeauthoryear{Van Der~Walt, Colbert  \& Varoquaux}{Van
  Der~Walt et~al.}{2011}]{van2011numpy}
Van Der~Walt S.,  Colbert S.~C.,   Varoquaux G.,  2011, Computing in Science \&
  Engineering, 13, 22

\bibitem[\protect\citeauthoryear{{Viero} et~al.,}{{Viero}
  et~al.}{2013}]{2013ApJ...772...77V}
{Viero} M.~P.,  et~al., 2013, \mn@doi [\apj] {10.1088/0004-637X/772/1/77},
  \href {https://ui.adsabs.harvard.edu/abs/2013ApJ...772...77V} {772, 77}

\bibitem[\protect\citeauthoryear{{Virtanen} et~al.,}{{Virtanen}
  et~al.}{2020}]{2020SciPy-NMeth}
{Virtanen} P.,  et~al., 2020, \mn@doi [Nature Methods]
  {https://doi.org/10.1038/s41592-019-0686-2}, \href {https://rdcu.be/b08Wh}
  {17, 261}

\bibitem[\protect\citeauthoryear{{Wang} et~al.,}{{Wang}
  et~al.}{2015}]{2015MNRAS.449.4476W}
{Wang} L.,  et~al., 2015, \mn@doi [\mnras] {10.1093/mnras/stv559}, \href
  {https://ui.adsabs.harvard.edu/abs/2015MNRAS.449.4476W} {449, 4476}

\bibitem[\protect\citeauthoryear{{Yan} et~al.,}{{Yan}
  et~al.}{2021}]{2102.07701}
{Yan} Z.,  et~al., 2021, arXiv e-prints, \href
  {https://ui.adsabs.harvard.edu/abs/2021arXiv210207701Y} {p. arXiv:2102.07701}

\bibitem[\protect\citeauthoryear{{Yan}, {van Waerbeke}, {Wright}, {Bilicki},
  {Gu}, {Hildebrandt}, {Maniyar}  \& {Tr{\"o}ster}}{{Yan}
  et~al.}{2022}]{2022arXiv220401649Y}
{Yan} Z.,  {van Waerbeke} L.,  {Wright} A.~H.,  {Bilicki} M.,  {Gu} S.,
  {Hildebrandt} H.,  {Maniyar} A.~S.,   {Tr{\"o}ster} T.,  2022, arXiv
  e-prints, \href {https://ui.adsabs.harvard.edu/abs/2022arXiv220401649Y} {p.
  arXiv:2204.01649}

\bibitem[\protect\citeauthoryear{Zonca, Singer, Lenz, Reinecke, Rosset, Hivon
  \& Gorski}{Zonca et~al.}{2019}]{Zonca2019}
Zonca A.,  Singer L.,  Lenz D.,  Reinecke M.,  Rosset C.,  Hivon E.,   Gorski
  K.,  2019, \mn@doi [Journal of Open Source Software] {10.21105/joss.01298},
  4, 1298

\bibitem[\protect\citeauthoryear{{Zou} et~al.,}{{Zou}
  et~al.}{2019}]{1908.07099}
{Zou} H.,  et~al., 2019, \mn@doi [\apjs] {10.3847/1538-4365/ab48e8}, \href
  {https://ui.adsabs.harvard.edu/abs/2019ApJS..245....4Z} {245, 4}

\bibitem[\protect\citeauthoryear{{van Engelen}, {Bhattacharya}, {Sehgal},
  {Holder}, {Zahn}  \& {Nagai}}{{van Engelen}
  et~al.}{2014}]{2014ApJ...786...13V}
{van Engelen} A.,  {Bhattacharya} S.,  {Sehgal} N.,  {Holder} G.~P.,  {Zahn}
  O.,   {Nagai} D.,  2014, \mn@doi [\apj] {10.1088/0004-637X/786/1/13}, \href
  {https://ui.adsabs.harvard.edu/abs/2014ApJ...786...13V} {786, 13}

\makeatother
\end{thebibliography}

\appendix

\section{$\bsfr$ constraints from alternative analysis choices}\label{app:tables}
  Table \ref{tab:binsep} shows the constraints on $\bsfr$ obtained by treating each redshift bin independently. The constraints found with a less conservative scale cut of $k_{\rm max}=0.2\,{\rm Mpc}^{-1}$ are shown in Table \ref{tab:k=02}. Finally, the results found including only one CIB frequency map at a time are shown in Table \ref{tab:freqsep}. The correlation matrix for the measurements presented in the Table \ref{tab:k=02} is
  \begin{equation}
    {\sf r}=\left(
    \begin{array}{cccccc}
      1.00 & 0.11  & 0.02  & 0.03 & 0.00  & 0.01 \\
      0.11 & 1.00  & 0.28  & 0.04 & 0.01 & -0.01 \\
      0.02 & 0.28  & 1.00  & 0.25 & 0.00 & -0.01 \\
      0.03 & 0.04  & 0.25  & 1.00 & 0.01  & 0.00 \\
      0.00 & 0.01  & 0.00  & 0.01 & 1.00  & 0.05 \\
      0.01 & -0.01 & -0.01 & 0.00 & 0.05  & 1.00
    \end{array}
    \right).
  \end{equation}
  \begin{table}
    \centering
    \def\arraystretch{1.4}
    \begin{tabular}{|ccccc|}
      \hline
      \multirow{2}{*}{Bin} & 
      \multirow{2}{*}{$b_g$} & $\bsfr$ &
      \multirow{2}{*}{$\chi^{2}/N_{\rm d.o.f}$} & 
      \multirow{2}{*}{PTE} \\
      & & ($\usfr$) & &\\
      \hline
      1   & $1.073\pm 0.044$ & $-0.002\pm 0.008$ & $1.75/10 = 0.175$ & $0.998$ \\
      2   & $1.377\pm 0.026$ & $0.024\pm 0.006$ & $19.98/18 = 1.109$ & $0.334$ \\
      3   & $1.307\pm 0.017$ & $0.041\pm 0.005$ & $34.18/26 = 1.315$ & $0.131$ \\
      4   & $1.733\pm 0.019$ & $0.048\pm 0.005$ & $23.38/30 = 0.779$ & $0.799$ \\
      5   & $2.069\pm 0.146$ & $0.111\pm 0.013$ & $56.23/50 = 1.125$ & $0.253$ \\ 
      6   & $2.158\pm 0.192$ & $0.264\pm 0.032$ & $61.43/54 = 1.138$ & $0.227$ \\
      \hline
    \end{tabular}
    \caption{Summary of the result values for the galaxy bias and the bias weighted star-formation rate within each redshift bin as presented in \ref{fig:zbins}, when all parameters are sampled bin by bin.}\label{tab:binsep}
  \end{table} %
  \begin{table}
    \centering
    \def\arraystretch{1.4}
    \begin{tabular}{|ccccc|}
      \hline
      \multirow{2}{*}{Bin} & 
      \multirow{2}{*}{$b_g$} & $\bsfr$ \\
      & & ($\usfr$)\\
      \hline
      1   & $1.125\pm 0.030$ & $0.012\pm 0.005$ \\
      2   & $1.366\pm 0.018$ & $0.023\pm 0.004$ \\
      3   & $1.307\pm 0.012$ & $0.042\pm 0.004$ \\
      4   & $1.739\pm 0.014$ & $0.053\pm 0.004$ \\
      5   & $1.970\pm 0.135$ & $0.118\pm 0.013$ \\ 
      6   & $2.321\pm 0.152$ & $0.271\pm 0.024$ \\
      \hline
      \multicolumn{3}{c|}{$\chi^{2}/N_{\rm d.o.f}=270.1/232 = 1.16, \hspace{12pt}{\rm PTE}=0.044$}\\
      \hline
    \end{tabular}
    \caption{Summary of the result values for the galaxy bias and the bias weighted star-formation rate for the least restrictive value of $k_{\rm max}=0.20\,{\rm Mpc}^{-1}$.}\label{tab:k=02}
  \end{table} %
  \begin{table}
    \centering
    \def\arraystretch{1.4}
    \begin{tabular}{|ccc|}
      \multirow{2}{*}{Bin} & 
      \multirow{2}{*}{$b_g$} & $\bsfr$ \\
      & & ($\usfr$)\\
      \hline
      \multicolumn{3}{c|}{$353$ GHz} \\
      \hline
      1   & $1.071\pm 0.045$ & $0.008\pm 0.025$ \\
      2   & $1.382\pm 0.026$ & $0.044\pm 0.017$ \\
      3   & $1.312\pm 0.016$ & $0.045\pm 0.014$ \\
      4   & $1.736\pm 0.019$ & $0.063\pm 0.012$ \\
      5   & $2.064\pm 0.149$ & $0.133\pm 0.020$ \\
      6   & $2.244\pm 0.186$ & $0.263\pm 0.033$ \\
      \hline
      \multicolumn{3}{c|}{$\chi^{2}/N_{\rm d.o.f}=84.6/88 = 0.96, \hspace{12pt}{\rm PTE}=0.583$}\\
      \hline
      \hline
      \multicolumn{3}{c|}{$545$ GHz} \\
      \hline
      1   & $1.073\pm 0.045$ & $0.000\pm 0.017$ \\
      2   & $1.382\pm 0.026$ & $0.041\pm 0.012$ \\
      3   & $1.311\pm 0.017$ & $0.036\pm 0.010$ \\
      4   & $1.737\pm 0.019$ & $0.054\pm 0.009$ \\
      5   & $2.062\pm 0.151$ & $0.128\pm 0.017$ \\
      6   & $2.244\pm 0.187$ & $0.254\pm 0.030$ \\
      \hline
      \multicolumn{3}{c|}{$\chi^{2}/N_{\rm d.o.f}=89.5/88 = 1.02, \hspace{12pt}{\rm PTE}=0.440$}\\
      \hline
      \hline
      \multicolumn{3}{c|}{$857$ GHz} \\
      \hline
      1   & $1.073\pm 0.045$ & $-0.003\pm 0.011$ \\
      2   & $1.382\pm 0.026$ & $0.034\pm 0.008$ \\
      3   & $1.310\pm 0.017$ & $0.039\pm 0.007$ \\
      4   & $1.736\pm 0.019$ & $0.051\pm 0.007$ \\
      5   & $2.075\pm 0.148$ & $0.119\pm 0.014$ \\
      6   & $2.246\pm 0.185$ & $0.227\pm 0.030$ \\
      \hline
      \multicolumn{3}{c|}{$\chi^{2}/N_{\rm d.o.f}=94.7/88 = 1.08, \hspace{12pt}{\rm PTE}=0.296$}\\
      \hline
    \end{tabular}
    \caption{Summary of the result values for the galaxy bias and the bias weighted star-formation rate within each redshift bin as presented in \ref{fig:zbins}, when all parameters are sampled frequency per frequency.}\label{tab:freqsep}
  \end{table} %

\bsp	
\label{lastpage}
\end{document}